\documentclass{iopart}
\usepackage{iopams}
\usepackage{graphicx}
\usepackage{color}

\newcommand{\ve}[1]{{\mathbf #1}}
\newcommand{\up}{\uparrow}
\newcommand{\down}{\downarrow}
\newcommand{\bk}{{\bf k}}

\newcommand{\bq}{{\bf q}}
\renewcommand{\k}{{\bf k}}
\newcommand{\p}{{\bf p}}

\newcommand{\0}{{\bf 0}}
\newcommand{\bP}{{\bf P}}

\newcommand{\ket}[1]{\left|{#1}\right.\rangle}
\newcommand{\ef}{\epsilon_F}

\newcommand{\beq}{\begin{equation}}
\newcommand{\eeq}{\end{equation}}
\newcommand{\fcal}{{\mathcal F}}

\begin{document}

\title[Polarons, Dressed Molecules, and Itinerant Ferromagnetism]
{Polarons, Dressed Molecules, and Itinerant Ferromagnetism in ultracold Fermi gases}

\author{Pietro Massignan}
\address{ICFO -- Institut de Ci\`encies Fot\`oniques, Mediterranean Technology Park, E-08860 Castelldefels (Barcelona), Spain}
\ead{pietro.massignan@icfo.es}

\author{Matteo Zaccanti}
\address{LENS and Dipartimento di Fisica e Astronomia, Universit\`a di Firenze,
  and INO-CNR, I-50019 Sesto Fiorentino, Italy}
\ead{zaccanti@lens.unifi.it}

\author{Georg M. Bruun}
\address{Department of Physics and Astronomy, University of Aarhus, Ny Munkegade, DK-8000 Aarhus C, Denmark}
\ead{bruungmb@phys.au.dk}

\begin{abstract}
In this review, we discuss the properties of a few impurity atoms immersed in a gas of ultracold fermions,  the so-called Fermi polaron problem.
On one side, this many-body system is appealing because it can be described almost exactly with simple diagrammatic and/or variational theoretical approaches. On the other, it provides quantitatively reliable insight into the phase diagram of strongly interacting population imbalanced quantum mixtures.
In particular, we show that the polaron problem can be applied to  study  itinerant ferromagnetism, a long standing  problem in quantum mechanics.
\end{abstract}

\pacs{05.30.Fk, 67.85.-d, 71.38.-k, 75.10.Lp}

\date{\today}

\tableofcontents

\section{Introduction}

A major challenge of modern physics is the study of quantum matter, aiming at understanding the wealth of phases realized by large ensembles of interacting particles at  low temperatures, where quantum mechanics plays a key role in determining their properties.
Quantum matter includes systems spanning an enormous range of energies, from ultracold gases and liquid Helium to electrons in solids,
all the way up to nuclear matter and quark-gluon plasmas.
Importantly, a common set of ideas and technical tools can be applied to these seemingly different systems, the investigation of one system yielding  information on another.
For instance, resonantly interacting cold atomic gases
and nuclear matter  are both essentially non-relativistic quantum matter with a scattering length much larger than the interparticle spacing, which has recently resulted in much cross-fertilization between the two fields~\cite{Hammer2013,Adams2013}.

Ubiquitous in quantum physics is the study of the so-called ``impurity problem", i.e.\ the investigation of the properties of a few particles immersed in a complex environment.
In their seminal works, Landau and Pekar proposed that the properties of conduction electrons in a dielectric medium could be understood in terms of so-called {\it polarons},
 i.e., quasi-particles resulting from the dressing of the electrons by collective excitations in the material \cite{Landau1933,Pekar1946}. This innovative idea was further elaborated by Fr\"ohlich and Feynman, who treated the ionic crystal or polar semiconductor as a phonon bath \cite{Frohlich1950,Frohlich1954,Feynman1955,Mahan2000book}.
  Other celebrated examples are the studies of
 helium-3 impurities in a bosonic helium-4 bath, and of the Kondo effect caused by localized magnetic impurities in metals.

The realisation of polaron physics in ultracold atomic gases has lead to a dramatic increase of activity in this topic.
 In these systems, polarons are realized by means of a population-imbalanced atomic gas,
  the minority atoms playing the role of \textit{impurities}, and the  majority atoms playing the role of the bath, or \textit{medium}.
The properties of the system depend strongly on the quantum nature of the bath. Experimentally, this scenario has been realized with a medium composed either of bosonic~\cite{Chikkatur2000,Catani2012,Spethmann2012,Scelle2013} or fermionic~\cite{Schirotzek2009,Kohstall2012,Koschorreck2012} atoms.
The case of impurities in a Bose gas may be viewed as an analog of the Fr\"ohlich polaron~\cite{Cucchietti2006,Tempere2009,Rath2013}: in the context of ultracold atoms, the change in the dynamical properties of the impurity is due to its interaction with excitations in the Bose gas, which are Bogoliubov ones in a condensate, or single-particle ones in a thermal gas.
 The other case, in which the medium is a low temperature fermionic gas, results instead in excitations named Fermi polarons, which are a paradigmatic realization of Landau's fundamental concept of a quasiparticle. The Fermi polaron is a main focus of this review, and we will refer to this fermionic case with the terms polaron and ``impurity problem" in the following.

An unexpected result  of the cold atom studies is that most of the polaron properties can be
 accurately described in the strong coupling regime using a  theory which is  much simpler than the ones required for a quantitative study of $50-50$ balanced fermi mixtures.
 Simple and quantitatively accurate theories for strongly interacting many-body systems are  rare, and the polaron problem therefore provides an
 important benchmark and starting point for improving our understanding of other strongly correlated systems. Notably, it turns out that the study of the single impurity case provides accurate information for strongly interacting polarized gases, even in the case of a sizeable concentration of minority particles. The behavior of these systems is governed by a simple equation of state written in terms of weakly-interacting quasiparticles in the spirit of Landau Fermi liquid theory.

The other main topic of this review is {\it itinerant ferromagnetism}, which is a kind of magnetism where the microscopic magnetic moments are mobile rather than localized.
 Itinerant ferromagnetism plays a crucial role for a large variety of metals \cite{Belitz2012} and other systems such as quark liquids in neutron stars \cite{Tatsumi2000}, and it
  has been intensely investigated ever since its theoretical prediction  shortly after the development of quantum mechanics~\cite{Stoner1933}.
 Itinerant ferromagnetism is however notoriously difficult to understand:
 It is in fact  still debated whether a homogenous electron system becomes ferromagnetic at all.
Moreover, the inevitable presence of disorder and of intricated band structures in solid state systems greatly complicates the comparison between microscopic theories and experiments.
 As we will discuss in this review, atomic gases can cast new light on this problem since one can investigate itinerant ferromagnetism in regimes which have been impossible to reach using other systems, and where the accurate polaron theory is applicable.
 There are however severe complications, both at a conceptual and practical level, arising from the short lifetime of a gas with a strongly repulsive interactions arising from short-range attractive potentials, which first have to be overcome.
As a consequence, the investigation of itinerant ferromagnetism with ultracold gases is still at a relatively early stage, both from the theoretical and from the experimental side.

Extensive reviews concerning  both balanced and polarized Fermi gases are  available in
literature~\cite{Varenna2007,Gurarie2007,Giorgini2008,Chevy2010,Gubbels2013,Randeria2013}.
With this work, we focus on  the recent experimental and theoretical advances in the understanding of the strongly polarized case
with focus on the repulsive interactions. We begin, in Sec.~\ref{sec:BasicScenario}, by reviewing the effective interaction between atoms at low energies, including effects due to a large effective range in the scattering amplitude. We also discuss the basic properties of one impurity (or very few ones) interacting with a large ideal Fermi gas, and we introduce the distinct quasiparticles which arise in this system.
In Sec.~\ref{sec:ManyBody} we analyze in detail the properties of the quasiparticles, and we present the various theories available and their surprising accuracy, as well as their shortcomings.
In the second part of the review, contained in Sec.~\ref{sec:IFM}, we introduce the reader to the phenomenon of itinerant ferromagnetism in the context of ultracold Fermi gases. We review the variety of theoretical approaches which have been employed for the study of a two fermion mixture with repulsive interactions, distinguishing the cases of purely repulsive potentials and short-range potentials. In this framework, we show how the knowledge of the repulsive polaron features allows one to obtain the phase diagram of a repulsively interacting two-fermion mixture with a small concentration of impurity atoms. In particular, we discuss how this
simple yet very accurate theoretical approach provides a deep insight into the competition between ferromagnetism and the decay and instability towards pairing, which is intrinsic for the repulsive state of ultracold Fermi gases. A comprehensive analysis of recent experiments with ultracold atoms is also presented. We conclude in Sec.~\ref{sec:Conclusions}, presenting a short summary and discussing interesting further developments and future research directions.


\section{Basic scenario and theoretical methods}
\label{sec:BasicScenario}
In this section, we  introduce the basic tools which allow for the investigation of the impurity problem in the contest of ultracold Fermi gases.
We start by recalling some basic results of scattering theory, discussing the effective interaction between the atoms in the ultracold regime.
Many excellent textbooks and reviews describe extensively ultracold collisions and Feshbach resonances~\cite{Landau1977book,Gurarie2007,PethickSmith2008book,Chin2010}, so we will focus only on the aspects which are most relevant for the following sections.
 Then we turn to the main topic of this review, i.e.\ the determination of the various states realised by a single minority
atom, denoted $\downarrow$, interacting with an ideal gas of majority particles, denoted $\uparrow$.
We first describe a simple model, which nevertheless captures most of the important features of the real $N_{\uparrow}+1_{\downarrow}$ system. Then the problem is analyzed by perturbation theory, and using an expansion in
the number of particle-hole excitations in the Fermi sea. Finally, we discuss various Monte-Carlo calculations, and we present the theoretical framework needed to describe the main experimental probe of the impurity properties, i.e., RF spectroscopy.

\subsection{Effective atom-atom interaction}
\label{scattering}
The description of interaction effects in ultracold atomic gases is simplified greatly by two effects.
First, the typical interparticle distance is much larger than the range of the inter-particle potential, which is of order of the van der Waals length
$r_{\rm vdW}\sim 100a_0$, with $a_0$ the Bohr radius. This  ensures
that only pairwise interactions must be taken into account.
Second, the ultralow temperatures combined with the short range character of the interatomic potential means that generally there is only appreciable
scattering in the $s$-wave channel.
Similarly to a light wave scattering from an object that is much smaller than its wavelength, the long de Broglie wavelength (up to micron size) associated to the motion of two ultracold atoms colliding via a short range potential produces predominantly isotropic, i.e. $s$-wave, scattering. As a consequence, the identical $\up$ fermions may be safely considered as ideal. Interactions are present only between the impurity $\down$ and the Fermi sea of $\up$ particles.

 The scattering wave function for the  $\uparrow$-$\downarrow$ atom pair outside of the scattering region reads
 $\psi({\mathbf r})=\exp(ikz)+f(k)\exp(ikr)/r$, where $f(k)$ is the scattering amplitude in the $s$-wave channel and $k$ is the relative momentum.
 At small momenta, the scattering amplitude in vacuum may be written as 
\begin{equation}
f_{\rm vac}(k)=-\frac{1}{k\cot\delta+ik}\approx-\frac{1}{a^{-1}-r_{\rm eff}k^2/2+ik }.
\label{ScattAmp}
\end{equation}
This expansion, which exhibits a Breit-Wigner resonance shape, contains only two parameters, namely the scattering length $a$ and the effective range $r_{\rm eff}$. Importantly, while the particular values of $a$ and $r_{\rm eff}$ are set by the microscopic details of the  interatomic potential,
completely different microscopic interactions can lead to the same low-energy scattering amplitude. As a consequence, one can substitute the true complicated interatomic interaction with a much simpler effective interaction parametrized by the scattering length $a$ and the effective range $r_{\rm eff}$.

The  poles of the scattering amplitude determine the energy of the bound states of the interaction potential. Introducing the {\it range parameter} $R^*=-r_{\rm eff}/2$, and restricting ourselves to the case $R^{*}>0,$ we see that Eq. (\ref{ScattAmp}) admits a single pole for $a>0$, describing a dimer with binding energy
(we set $\hbar=1$ throughout in this review)
\beq
 E_{\rm d}^{\rm vac}=-1/2m_r a_*^2
\label{dimerEnergyInVacuum}
\eeq
 with $a_*=2R^*/(\sqrt{1+4R^*/a}-1)$, and $m_r=m_\up m_\down/(m_\up + m_\down)$ the reduced mass of the two particles.
For $R^{*}\ll a$, we recover the \textit{universal} dimer energy $E_{\rm d}^{\rm vac}=-1/2m_ra^2$, whereas we get $E_{\rm d}^{\rm vac}=-1/2m_rR^{*}a$ when $R^{*}\gg a$.

 Even though in this review we allow the masses of the $\up$ and $\down$ particles to differ, we will restrict ourselves to small enough mass ratios, i.e. $m_\up<13.6m_\downarrow$ to avoid the three-body Efimov effect, and even $m_\up<13.384m_\downarrow$ to avoid the four-body Efimov effect. In this regime, the results are independent of the short-distance behavior of the interatomic potential \cite{Braaten2006,Castin2010}. Under this condition, the bare interaction between particles may be taken as momentum-independent, or in other words, one can replace the bare two-body potential by a regularized pseudo-potential of zero range (or contact potential). Unless otherwise specified, results presented in this review are obtained in this limit\footnote{A model retaining explicitly a finite interaction range of the potential is discussed in detail in \cite{Werner2006}}.
 
In order to include many-body effects, it is often simpler to work in momentum space.
Solving the Schr\"odinger equation for the scattering wave function is equivalent to solving the  Lippmann-Schwinger equation, which is a matrix
equation for the   $\mathcal T$-matrix   in the  scattering channels consisting of the different atomic
hyperfine states coupled by the interaction~\cite{PethickSmith2008book}.
  In the simplest case, particles enter, collide, and leave the scattering region through a single channel. Within this ``one-channel model", the
   Lippmann-Schwinger equation is simply a scalar one with the solution
\beq
\mathcal{T}_{\rm 1ch}(\bP,\omega)=\frac{2\pi}{m_r}\left[\frac{1}{a}-\frac{2\pi\Pi(\bP,\omega)}{m_r}\right]^{-1}
\label{SingleChannel}
\eeq
for two particles with total momentum $\bP$ and energy $\omega$.
We have here introduced the renormalized propagator of a pair of free atoms in the medium (or pair propagator),
\beq
\Pi(\bP,\omega)=\int\frac{{\rm d}^3p}{(2\pi)^3}\left[\frac{1-f_\uparrow(\p)-f_\downarrow(\bP+\p)}{\omega+i0^{+}-\xi_{\uparrow\p}-\xi_{\downarrow\bP+\p}}+\frac{2m_r}{p^2}\right].
\label{pairPropagator}
\eeq
Medium effects are taken into account by means of the Fermi functions $f_\sigma(\p)=[\exp(\beta\xi_{\sigma\p})+1]^{-1}$, with
$\beta=1/k_BT$ the inverse temperature and $\xi_{\sigma\p}=p^2/2m_\sigma-\mu_\sigma$ the kinetic energy measured from the chemical potential $\mu_\sigma$.
The Fermi functions reflect the blocking of available states due to
 the presence of other identical fermions. This is the minimal step towards including  many-body effects in the scattering,
and it has significant physical effects. Surprisingly, we shall  see that for the physics of interest here this many-body correction is sufficient to obtain quantitative agreement
with experimental and Monte-Carlo results even for strong interactions.

In a vacuum, the "on-shell" $\mathcal T$-matrix, obtained by setting the energy to  $\omega=P^2/2M+k^2/2m_r$ with $\k$ the relative momentum and $M=m_\uparrow+m_\downarrow$, is related to the scattering amplitude by the relation ${\mathcal T}(k)=-2\pi f(k)/m_r$. Using $\Pi(0,k^2/2m_r)=-im_rk/2\pi$ in a vacuum, i.e.\ setting the Fermi functions to zero in (\ref{pairPropagator}),
the one-channel T-matrix reduces to $2\pi/[m_r \mathcal{T}_{\rm 1ch,vac}(k)]=a^{-1}+ik$.
Comparing with Eq.\ (\ref{ScattAmp}), we see that this one channel model yields a vanishing effective range of the interaction.

\begin{figure}[ht]
\begin{center}
\includegraphics[width=0.6\linewidth]{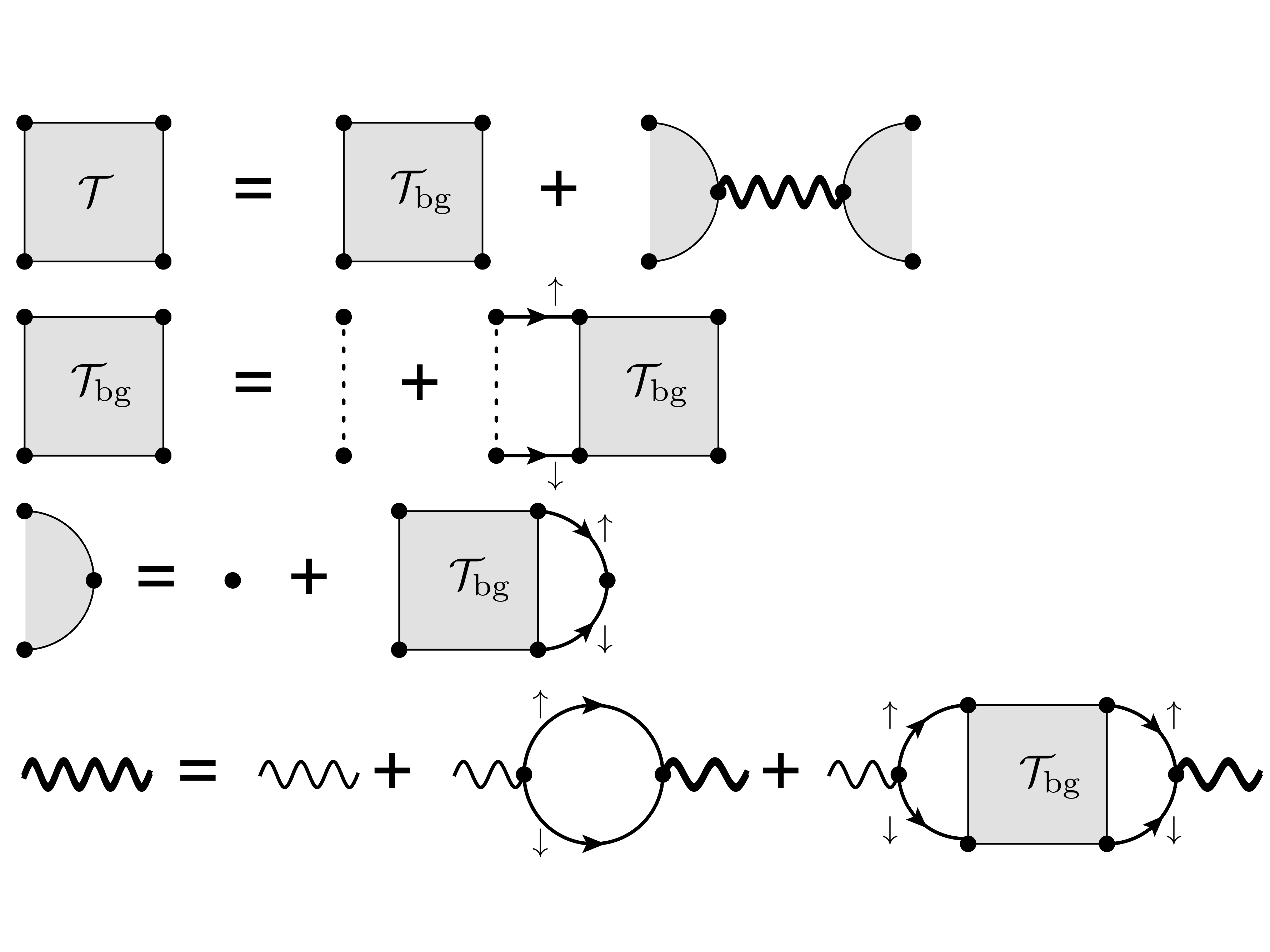}
\end{center}
\caption{The ${\mathcal T}$-matrix in the two-channel model within the ladder approximation. Lines are bare Green's functions for the open channel atoms, dashed lines represent the atom-atom interaction in the open channel, the thin wavy line represents the bare Feshbach molecule, and the thick wavy line is the 
Feshbach molecule including interactions with the open channel.}
\label{FeynmanTmatrix}
\end{figure}

A very powerful feature of atomic gases is that the scattering length can be tuned using a \emph{Feshbach resonance}~\cite{PethickSmith2008book,Chin2010}.
By tuning an external magnetic field, one can bring to degeneracy the energy of a pair of atoms in the scattering state (or open
 channel), with a bound molecular state supported by a different (closed) channel, provided the magnetic moment difference $\delta \mu$ between these two channels is nonzero. Whenever this happens, the scattering length diverges and the atomic gas becomes strongly interacting. To describe this, one needs a
  two-channel model for the scattering.
  Within a many-body diagrammatic approach, the $\mathcal T$-matrix describing incoming
  and outgoing particles in the open channel, which interact in the scattering region with the closed channel, can be written as \cite{Bruun2005,Massignan2008}
\beq
\frac{2\pi}{m_r \mathcal{T}(\bP,\omega)}=\frac{1}{\tilde{a}(E_{\rm CM})}-\frac{2\pi\Pi(\bP,\omega)}{m_r}.
\label{twoChannelTmatrix}
\eeq
We have here introduced an effective "energy-dependent scattering length"
\beq
\tilde a(E_{\rm CM})\equiv a_{\rm bg}\left(1-\frac{\Delta B}{B-B_0-E_{\rm CM}/\delta\mu}\right)
\label{aTilde}
\eeq
with $E_{\rm CM}=\omega-P^2/2M+\mu_\uparrow+\mu_\downarrow$ the energy in the center of  mass frame of the two particles. The Feynman diagrams 
corresponding to the two-channel model for the scattering  leading to Eq.\ (\ref{twoChannelTmatrix}), are shown in Fig.\ \ref{FeynmanTmatrix}.

With Eq.~(\ref{twoChannelTmatrix}), the scattering between a pair of  $\uparrow$ and $\downarrow$ atoms in a
medium is described completely in terms of the experimental parameters characterizing a Feshbach resonance: the resonance center $B_0$,  the resonance width $\Delta B$, the magnetic moment difference $\delta\mu$ between open and closed channels, and the background scattering length $a_{\rm bg}$.
The energy dependence of $\tilde a$ expresses the fact that the resonance energy is shifted due to the center of mass energy of the dimer.

A low energy expansion of Eq.~(\ref{twoChannelTmatrix}) in the vacuum case yields
\beq
\frac{2\pi}{m_r{\mathcal{T}(0,k^2/2m_r)}}=-\frac1{f(k)}\approx a^{-1}+ik+R^*k^2+\ldots
\eeq
with the familiar  behavior
\beq
a=a_{\rm bg}\left[1-\Delta B/(B-B_0)\right],
\eeq
and the range parameter
\beq
R^*=R^*_{\rm res}(1-a_{\rm bg}/a)^2
\eeq
with $R^*_{\rm res}=1/2m_ra_{\rm bg}\Delta B\delta\mu$  the value at resonance~\cite{Petrov2004,Bruun2005}.
 Thus, the two-channel model naturally yields a non-zero effective range for the scattering.
 For most resonances of interest
$k_F |a_{\rm bg}|< 1$, and in the strongly-interacting regime anyway one always has $|a_{\rm bg}/a|\ll1$. To simplify the treatment in the rest of the review,
we will work in this limit, where  $R^*=R^*_{\rm res}$.

At the two-body level, a resonance whose range parameter $R^*$ is small compared to $r_{\rm vdW}$
is termed {\it broad} \cite{Chin2010}. Since $|a|$ is generally at least of order $r_{\rm vdW}$, at a broad resonance the physics is universal, in the sense that $R^*$
plays a negligible role. A resonance for which
 $R^*$ is instead large compared to $r_{\rm vdW}$ is termed {\it narrow}.
However, in the context of strongly-interacting ultracold fermions, it is useful to employ an alternative criterion: since particles collide with typical momenta of order the Fermi momentum $k_F$, the correction due to the range parameter in the denominator of the scattering amplitude becomes negligible when $R^* \ll 1/k_F$. In the following, we will define a resonance to be broad if $k_F R^*\ll1$, and narrow if $k_F R^*\gg1$ \cite{Gurarie2007}.
For a broad resonance, Eq.\ (\ref{twoChannelTmatrix}) reduces to the usual single-channel scattering matrix (\ref{SingleChannel}).

\subsection{Energy spectrum for the $N_{\uparrow}+1_{\downarrow}$-body problem}\label{BasicPhysSec}
Having analyzed the effective interaction between the atoms, we  now focus on the excitations displayed by a fermionic  mixture where one $\downarrow$
atom interacts with a gas of $\uparrow$ atoms. The dimensionless parameter $k_Fa$, with $n_\up=k_F^3/6\pi^2$ the density of the majority atoms,
gives the interaction strength of this problem and the natural unit of energy  is the majority
Fermi energy $\epsilon_F=k_F^2/2m_\uparrow$. The strongly interacting regime corresponds to
 $1/k_F|a|\lesssim 1$, while we refer to $1/k_Fa\ll -1$ and  $1/k_Fa\gg 1$ as the BCS and BEC limits respectively.  Finally, we will measure energies relative to the
 non-interacting $N_{\uparrow}+1_{\downarrow}$ particle system, and we restrict ourselves to the thermodynamic limit $N_\up\gg 1$.

\subsubsection{A simple toy model}
\label{subsec:toyModel}
Before analysing the full many-body case, it is instructive to consider a simple toy model, which displays in an intuitive way many of the features of the impurity problem\footnote{This model was first introduced in Refs.~\cite{Pricoupenko2004,CastinVarenna2007} for a balanced (50-50) mixture of $\up$ and $\down$ atoms, and we adapt it here
to the case of a single $\down$ minority in a Fermi gas of spin $\up$ particles, and to the case of a non-zero range parameter.}.
The great simplification of this model relies on the assumption that the $\down$ atom interacts only with the closest $\up$ atom, while the presence of
 the rest of the $\up$ atoms in the Fermi sea is accounted for by means of a  boundary condition. In the center of mass frame
 of the  $\up\down$ pair, the two-body problem reduces to a single particle of reduced mass $m_r=m_\up m_\down/(m_\up+m_\down)$ scattering
  on an infinitely massive object located at $r=0$.
    The short range potential at the origin is described by the Bethe-Peierls boundary condition for
  the relative wavefunction $\psi$
  \beq
\lim_{r\rightarrow 0}\frac{\partial_r(r\psi)}{(r\psi)}=k\cot(\delta)=-\frac1 a-R^*k^2,
\label{BethePeierlsBoundaryConditions}
\eeq
where $\delta$ is the phase shift, and $\psi=\sin(kr+\delta)/r$ is the positive energy $\epsilon=k^2/2m_r$ solution to the  Schr\"odinger equation
\beq
-\frac{\nabla^2\psi}{2m_r}=\epsilon\psi
\label{freeSchroedingerEquation}
\eeq
for $r>0$. The presence of the remaining $\up$ particles acts as a Fermi pressure on the $\up$ particle.
 We model this by requiring  that the wavefunction $\psi$ vanishes at the boundary of a spherical cavity of radius $R$, i.e.\ we use the boundary condition
  $\delta=-kR$. Equating the
 ground state energy in absence of interaction, $(\pi/R)^2/2m_r$, with
  the energy $\epsilon_F$ needed to add a pair of  $\up\downarrow$ particles to the  Fermi sea, one finds
\beq
R=\frac{\pi}{k_F}\sqrt{\frac{m_\up}{m_r}},
\eeq
i.e., a radius of order the average interparticle distance. The boundary conditions at the origin and at the edge of the box  then yield the relation $k\cot (kR)=a^{-1}+R^{*}k^{2}$.
 For negative energy $\epsilon=-\kappa^2/2m_r$, correspondingly one has $\psi=\sinh[\kappa(r-R)]/r$
  and the condition $\kappa\coth (\kappa R)=a^{-1}-R^{*}\kappa^{2}$.

\begin{figure}
\begin{center}\includegraphics[width=0.6\linewidth]{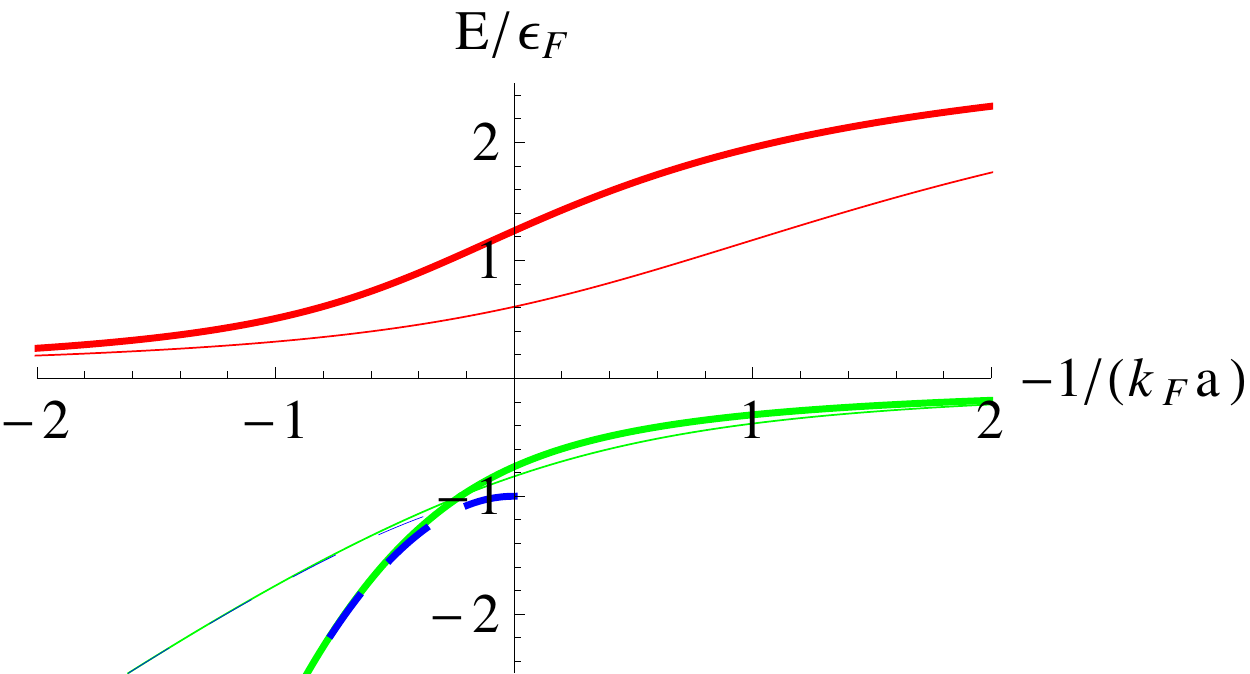}\end{center}
\caption{Energy of the two lowest branches of the toy model for the case $m_\down=m_\up$. We plot $E=\epsilon-\epsilon_F$, such that $E$ tends to zero in the non-interacting limit $|k_Fa|\ll 1$. Thick lines represent a broad resonance, while thin ones a narrow one with $k_FR^{*}=1$. The dashed blue lines are the vacuum dimer energies $E_{\rm d}^{\rm vac}-\epsilon_F$.
}
\label{fig:toyModel}
\end{figure}

In Fig.\ \ref{fig:toyModel} we plot the  two lowest energy solutions, or branches, of this model for the equal mass case $m_\up=m_\down$. The two branches continuously
connect  in the non-interacting limit $a=0$. To first order in $k_Fa$, the energy shift of the upper (lower) branch increases (decreases) as $3(m_r/m_\up)^{3/2}[2\pi an_\up/m_r]$,
 with the factor in square brackets equal to the mean field shift. For equal masses one finds $3(m_r/m_\up)^{3/2}=1.061$. The toy model therefore
 almost reproduces the exact result in the weak coupling limit.
In the vicinity of the resonance, the scattering length is much larger than the average interparticle distance and it drops out of the problem: the energy of the two branches then depends solely on the Fermi energy of the system, and on the range
parameter $R^*$.
On the BEC side $a>0$,  the attractive polaron energy approaches  the vacuum dimer energy $E_{d}^{\rm vac}-\epsilon_F$, reflecting the fact that the ground state for $a\rightarrow 0^+$
is a  tightly-bound $\uparrow\downarrow$ dimer immersed in the majority Fermi sea.

This simple model captures some of the essential physics of the impurity problem, but being a two-body approximation to a many-body problem one cannot
 hope to obtain quantitatively correct results. Furthermore, the model misses two qualitative features which play a key role in the rest of this review:
 the metastability of the upper branch, and the presence of a dressed dimer+hole continuum. Nonetheless, this toy model gives a simple physical picture which demonstrates the presence of a repulsive branch of excitations in a system with purely attractive interactions.

\begin{figure}[ht]
\begin{center}\includegraphics[width=0.8\linewidth]{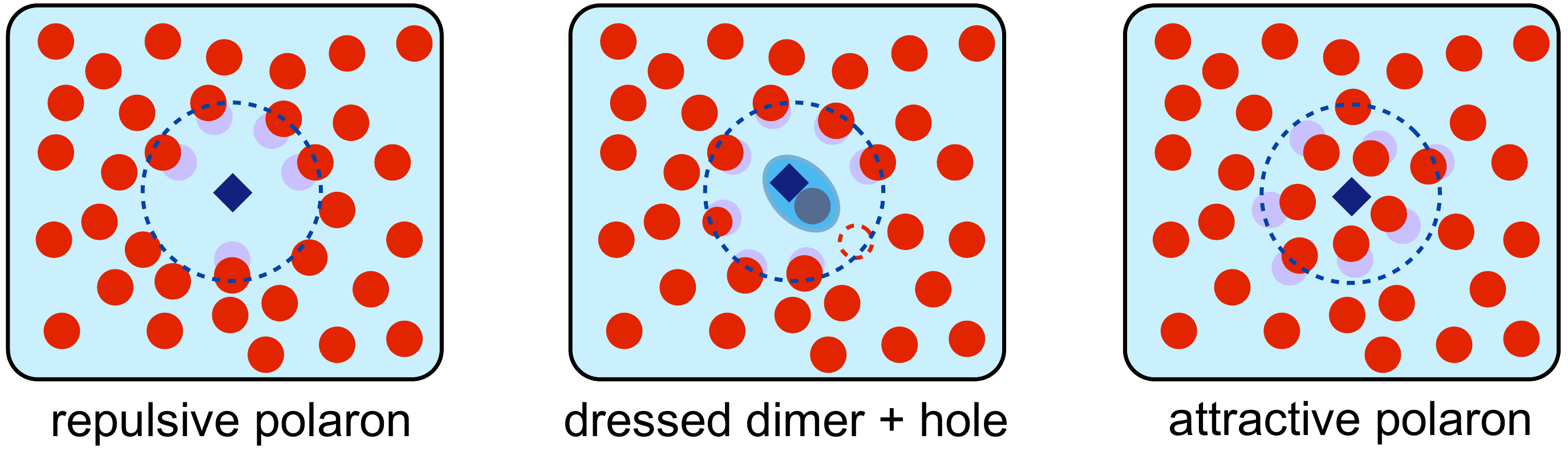}\end{center}
\caption{Possible states of an impurity immersed in a Fermi gas.}
\label{fig:sketchminorityStates}
\end{figure}

\begin{figure}
\includegraphics[width=0.6\linewidth]{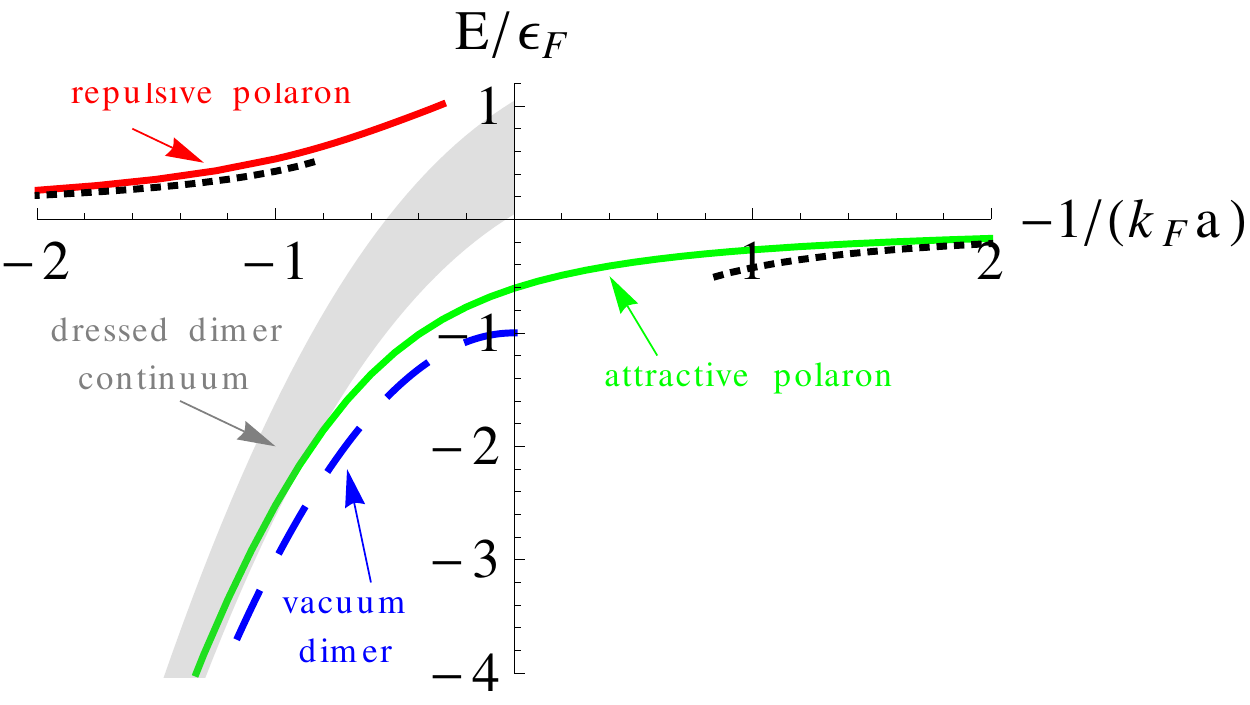}
\caption{The energy spectrum of a zero-momentum impurity in a Fermi sea contains a repulsive polaron, a continuum of dressed dimers, and an attractive polaron. The dotted black lines are the mean-field result, and the dashed line is the dimer energy in absence of the Fermi sea. The spectrum is generic but the quantitative details correspond here to the case $m_\up=m_\down$ and $R^*=0$. The curves are obtained from the 1PH approximation described in Sec.~\ref{ManybodyModels}.
}
\label{fig:minoritySpectrum}
\end{figure}

\subsubsection{Complete many-body picture}
\label{subsec:completeManyBodyPicture}
Before entering into the detail of their derivation, we present here  the complete set of states present in the spectrum of one single minority atom at zero momentum embedded in a Fermi gas.
The possible states are showed pictorially in Fig.\ \ref{fig:sketchminorityStates}, while their energies are shown in Fig.\ \ref{fig:minoritySpectrum}.

 First, the lower green line in Fig.\ \ref{fig:minoritySpectrum} gives the energy  $E_-$ of a quasiparticle formed by the minority atom attracting a cloud of surrounding
majority atoms. It is referred to as the \textit{attractive polaron} in analogy with an electron interacting
with phonons in  a crystal~\cite{Mahan2000book}.
The attractive polaron has the energy $E_-=2\pi an_\up/m_r$ in the BCS limit, and it gets increasingly  bound with increasing $1/k_Fa$. At  resonance, $1/k_Fa=0$,
 for equal masses one finds $E_-\sim-0.6\epsilon_F$ when $k_FR^*\ll 1$.

Second, the upper red line in Fig.~\ref{fig:minoritySpectrum} gives the energy $E_+$ of the \textit{repulsive polaron}, a quasiparticle formed by the minority atom repelling the surrounding majority atoms.
The repulsive polaron  is the many-body analogue of the first excited scattering state of the two-body problem (cfr.\ the red line in Fig.\ \ref{fig:toyModel}). In the BEC limit,
  the energy is $E_+=2\pi an_\up/m_r$ and it increases to become of the order of $\epsilon_F$ as the  Feshbach resonance is approached from the BEC side. However,
the repulsive polaron becomes increasingly unstable towards decay (see Sec. \ref{subsec:decay})
as the resonance is approached from the BEC side, and it eventually becomes ill-defined.

Third, there is a continuum of states between the repulsive and attractive polaron energies in
Fig.~\ref{fig:minoritySpectrum}. In the BEC limit $1/k_Fa\gg 1$,
it consists of a dressed dimer+hole continuum, where the dressed dimer is formed by the minority atom and one majority atom, which can be taken from anywhere inside the Fermi sea, resulting in a  continuum of excitations with spectral width $\sim \epsilon_F$.

 The ground state of the mixture is a polaron on the BCS side,
 and a dressed dimer immersed in a Fermi sea with one less particle on the BEC side. When the polaron is the ground state, there is also a continuum 
 of zero momentum states corresponding to a finite momentum polaron and a particle-hole excitation in the Fermi sea with the opposite momentum, 
 which  extends all the way down to the zero momentum polaron energy~\cite{Trefzger2012}.

When the mass of the minority $m_\down$ is smaller than $m_\up/6.7$, the ground state of the mixture may even be a trimer \cite{Mathy2011}. Tetramers (and probably even larger compounds) generally exist in vacuum for mass imbalanced systems \cite{Castin2010,Levinsen2013,Blume2012}, but are unfavored in the presence of a Fermi sea \cite{Parish2013}.

\subsection{Many-body calculations}
\label{ManybodyModels}
In order to quantitatively study the polaron problem, one  needs a more sophisticated model than what was described in
Sec.~\ref{subsec:toyModel}. As we shall now discuss, one can in fact develop an accurate many-body theory which is
relatively simple.

Historically, the problem of one impurity in a Fermi gas was first studied by Bishop for a hard sphere interaction, which corresponds to
 a purely repulsive potential. A rigorous perturbative expansion in the interaction parameter $k_F a$ yields \begin{equation}
E_+=\epsilon_{F\uparrow}\frac{m_\uparrow}{m_r}\left[\frac{2}{3\pi}(k_Fa)+F(\alpha)(k_Fa)^2+O[(k_Fa)^3]\right],
\label{muDownAnalytic}
\end{equation}
for the  repulsive
 polaron energy~\cite{Bishop1973}, with $\alpha=(m_\downarrow-m_\uparrow)/(m_\downarrow+m_\uparrow)$ and
\begin{equation}
F(\alpha)=\frac{1-\alpha}{4\pi^2\alpha^2}\left[(1+\alpha)^2\log\left(\frac{1+\alpha}{1-\alpha}\right)-2\alpha\right].
\end{equation}
The first term in (\ref{muDownAnalytic})  is the mean-field result $E_+=2\pi a n_\up/m_r$.
For equal masses $m_\uparrow=m_\downarrow$ one finds
\beq
\frac{E_+}{\epsilon_F}=\frac{4 k_Fa}{3 \pi }
+\frac{2 (k_Fa)^2}{\pi ^2}
   +\left(\frac{4}{3}+\frac{2 \pi ^2}{45}\right)\frac{2  (k_Fa)^3}{\pi
   ^3}+\ldots.
   \label{BishopEnergy}
\eeq
 This perturbative result should of course be taken with caution in the strongly-interacting regime where $k_F|a|\gtrsim1$. It is also known that the third order term in the expansion is non-universal.\footnote{
Bishop also found an analytic expression for the effective mass of the minority particle:
\beq
\left(\frac{m^*_+}{m_\downarrow}\right)^{-1}=1-\frac{2}{3\pi^2}\left[\frac{1}{\alpha}-\frac{(1-\alpha)^2}{2\alpha^2}\log\left(\frac{1+\alpha}{1-\alpha}\right)\right](k_Fa)^2+\ldots
\eeq
 }

 Recently, the polaron has been analysed by means of a  variational Ansatz expanding the many-body wave function in terms of the number of particle-hole
 excitations in the  Fermi sea. The wave function is  written as~\cite{Chevy2006}
 \begin{equation}
|\psi\rangle= \sqrt{Z} a_{\ve{0}\downarrow}^\dagger|{\rm FS_N}\rangle +
\sum_{q<k_F<k}\phi_{\bq,\bk} a_{\ve{q}-\ve{k}\downarrow}^\dagger
a^{\dag}_{\ve{k}\uparrow}\,a_{\ve{q}\uparrow}|{\rm FS_N}\rangle + \ldots,
\label{polaronAnsatz}
\end{equation}
where $a_{{\bf k}\sigma}$ annihilates a fermion of species $\sigma$ with momentum ${\mathbf k}$,  $|{\rm FS_N}\rangle$ denotes the Fermi sea
of N $\uparrow$ particles,
and $\{Z,\phi_{\bq,\bk}\}$ are variational parameters.
In particular, the quantity $Z$, usually called \emph{quasiparticle residue}, represents the overlap between the quasiparticle and the bare particle states.
This variational function, first introduced to analyze the attractive polaron, was
later applied to investigate the repulsive polaron~\cite{Cui2010}, and to the case of  narrow resonances~\cite{Massignan2012,Qi2012,Trefzger2012}.

\begin{figure}[ht]
\begin{center}
\includegraphics[width=0.25\linewidth]{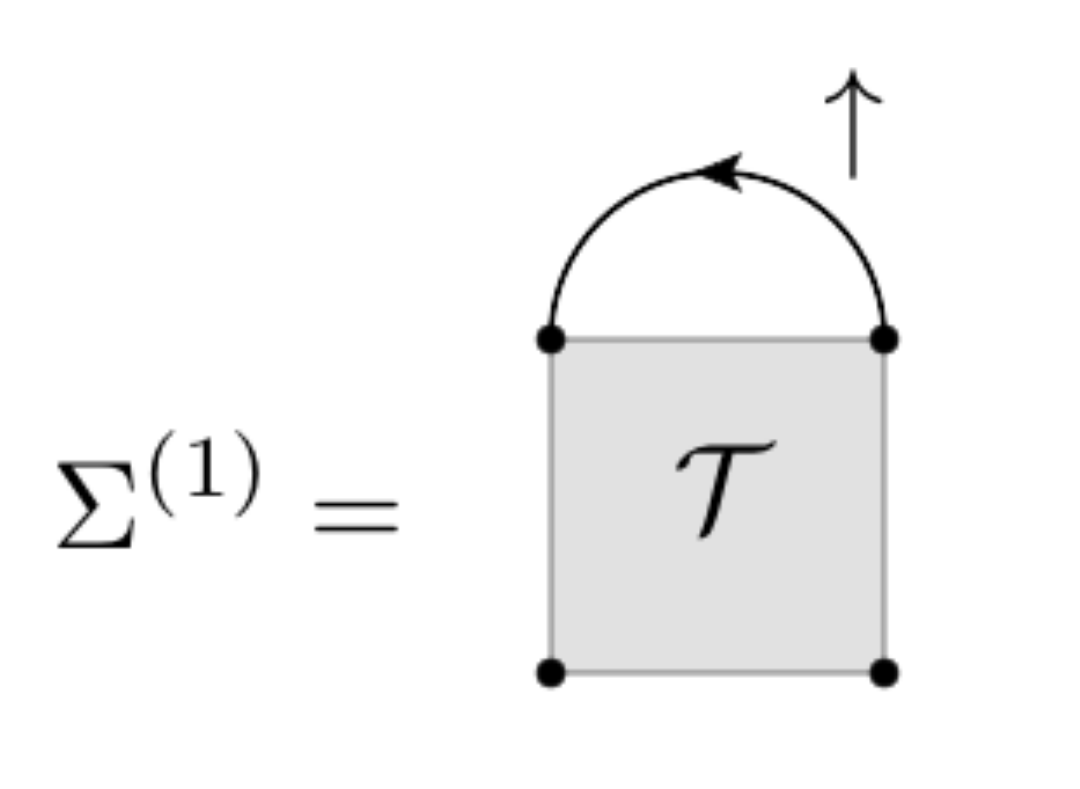}
\end{center}
\caption{The ladder approximation for the self-energy. }
\label{SelfEnergyFeyn}
\end{figure}

In order to describe both the energy and the decay of the polaron, it is natural to use diagrammatic
many-body theory. The Green's function of the impurity particle is
$G_\down(\p,\omega)=[\omega-p^2/2m_\down-\Sigma(\p,\omega)+i0_+]^{-1}$ where $\Sigma(p,\omega)$ is the self-energy.
The latter describes the energy shift of the minority particle due to the interactions with the medium.
The variational approach described above is equivalent to expanding the  self-energy $\Sigma(\p,\omega)$
 in the number of holes created in the  Fermi surface writing
\begin{equation}
\Sigma(\p,E)=\Sigma^{(1)}(\p,E)+\Sigma^{(2)}(\p,E)+\ldots
\label{SelfEnergy}
\end{equation}
where $\Sigma^{(n)}$ denotes processes involving $n$ holes. Truncating the series at the  one particle-hole level $\Sigma^{(1)}$ corresponds to  keeping the first two terms in (\ref{polaronAnsatz}). We will often refer in the following to this truncation as to the \emph{1PH approximation}. This gives the ladder
self-energy~\cite{Combescot2007,Massignan2011}
\begin{equation}
\Sigma^{(1)}(\p,E)= \int\frac{{\rm d}^3q}{(2\pi)^3}f_\uparrow(q){\mathcal{T}}({\mathbf q}+{\mathbf p},E+\xi_{q\uparrow}).
\label{Selfenergy}
\end{equation}
The Feynman diagram corresponding to Eq.~(\ref{Selfenergy}) is shown in Fig.~\ref{SelfEnergyFeyn}.

The energy of the repulsive and attractive polarons, shown respectively as red and green solid lines in Fig. \ref{fig:minoritySpectrum}, are found as solutions to the implicit equation
\beq
E_\pm={\rm Re}[\Sigma(0,E_\pm)],
\eeq
and the quasiparticle residue at a given pole is found from
\beq
Z_\pm=[1-\partial_\omega\Sigma(0,E)|_{E_\pm}]^{-1}.
\eeq
Setting $\Sigma(p,E)=\Sigma^{(1)}(p,E)$ yields the same value as the variational parameter $Z$ in (\ref{polaronAnsatz}). Also,
the effective mass of the polaron can be found as
\beq
m_\pm^*=\frac{m_\down}{Z_\pm}\left[1+\frac{\partial {\rm Re}[\Sigma(\p,E_\pm)]}{\partial(\epsilon_{p,\down})}\right]^{-1},
\eeq
and close to the polaron poles,  one may write the Green's function as
\begin{equation}
G_{\downarrow}(\ve{p},\omega)\sim\frac{Z_\pm}{\omega-E_\pm-p^2/2m_{\pm}^{*}+i0^+},
\label{polaronsPoleExpansion}
\end{equation}
explicitly showing that these quasiparticles may be thought of as free particles with renormalized masses and spectral weights.

One-particle-hole diagrammatic and variational treatments were subsequently developed also for the energy of the dressed dimers by \cite{Mora2009,Punk2009,Combescot2009}.
In close analogy with Eq.\ (\ref{polaronAnsatz}), the ansatz for the zero-momentum dressed dimer may be written as
 \begin{equation}
|\Psi\rangle=
\sum_{k>k_F}\phi_\bk a_{-\bk\down}^\dagger a_{\bk\up}^\dagger|{\rm FS_{N-1}}\rangle +
\sum_{q<k_F<k,k'}\phi_{\bq,\bk,\bk'}
a_{\bq-\bk-\bk'\down}^\dagger
a_{\bk\up}^\dagger
a_{\bk'\up}^\dagger
a_{\bq\uparrow}|{\rm FS_{N-1}}\rangle + \ldots,
\label{moleculeAnsatz}
\end{equation}
where $|\rm FS_{N-1}\rangle$ denotes a Fermi sea with $N-1$ $\up$ particles, and $\{\phi_\bk,\phi_{\bq,\bk,\bk'}\}$ are variational parameters. $N$-mers with $N>2$ (trimers,
tetramers, and so on) can be studied by suitable extensions of (\ref{polaronAnsatz}) and (\ref{moleculeAnsatz}).

The impurity problem has also been investigated by means of Quantum Monte-Carlo studies (QMC). For attractive interactions, \cite{Pilati2008} studied the case of a
finite density of impurities by means of Fixed-Node QMC, while \cite{Prokofev2008,Prokofev2008a,Vlietinck2013} looked at the case of a single minority with
bold-diagrammatic QMC, pointing out the presence of the dressed dimer+hole continuum. Later, the repulsive branch was analysed with QMC~\cite{Conduit2009,Pilati2010,Chang2011}.
Finally, the complete spectrum shown in Fig. \ref{fig:minoritySpectrum} was also calculated using the functional renormalization group \cite{Schmidt2011}.

When comparing the 1PH variational/diagrammatic results with the Monte-Carlo calculations,
one obtains a remarkable agreement for the energies of both the polaron and the dressed dimer,
even for strong interactions. For the attractive polaron, the agreement is at the 1\% level, whereas the agreement is slightly worse for the repulsive
polaron as we shall discuss later. However, the 1PH approximation is insufficient to describe the decay correctly as we will see.

\subsection{Radio-frequency spectroscopy and spectral function}
Radio-frequency (RF) spectroscopy has been applied very successfully in ultracold gases experiments to measure a variety of
properties, such as the size and temperature of an
atom cloud \cite{Martin1988,Bloch1999},  clock-shifts, creation of molecules, and pairing gaps~\cite{Chin2004,Shin2007,Stewart2008,Gaebler2010}. It is also  the main probe of the
polaron properties~\cite{Schirotzek2009,Kohstall2012,Koschorreck2012}, and we therefore now briefly review this method.

We consider RF  spectroscopy involving three spin states: $\ket \up$ for the majority, $\ket 0$, and $\ket 1$ for the minority component. The atoms are initially prepared in a
mixture of states $\ket \up$ and $\ket 0$.
By applying a weak RF field of suitable frequency, a small population of atoms in state $\ket 0$ is transferred to a third state $\ket 1$
which is initially empty.
The RF field is essentially uniform over the scale of the atomic cloud, and as such the photons do not transfer momentum to the atoms.
The action of the RF field may
then be described by the operator \cite{Massignan2008}
\begin{equation}
H_{\rm  rf}=\frac{\Omega}{2}\int d^3r\left[e^{-i\omega t}\psi_1^\dagger({\mathbf{r}},t)\psi_0({\mathbf{r}},t)+\rm{h.c.} \right],
 \label{RFTerm}
\end{equation}
where $\psi_i({\mathbf{r}},t)$ is the field operator for the atoms in state $|i\rangle$, $\Omega$ is the Rabi frequency describing
the coupling of the involved hyperfine states to the electromagnetic field, and $\omega$ is the rf frequency.
The induced transition rate $R(\omega)$ from state $|0\rangle$  to
$|1\rangle$ is within linear response (i.e., assuming that the population in state $|1\rangle$ remains negligible at all times) given by
\begin{equation}
R(\omega)\propto-{\rm Im}{\mathcal{D}}(\omega) \equiv -\int d^3rd^3r'{\rm Im}{\mathcal{D}}({\mathbf{r}},{\mathbf{r}}',\omega)
\label{Linresp}
\end{equation}
where ${\mathcal{D}}({\mathbf{r}},{\mathbf{r}}',\omega)$ is the Fourier transform of the retarded spin flip correlation function
$-i\theta(t-t')\langle[\psi_1^\dagger({\mathbf{r}},t)\psi_0({\mathbf{r}},t),\psi_0^\dagger ({\mathbf{r}}',t')\psi_1({\mathbf{r}}',t')]\rangle$.

\begin{figure}[ht]
\begin{center}
\includegraphics[width=0.49\linewidth]{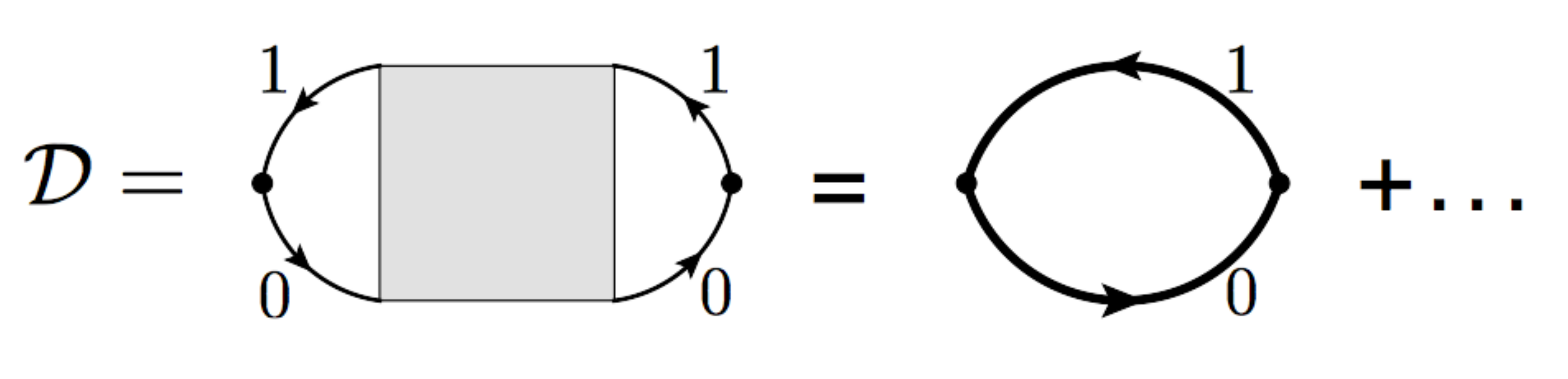}
\end{center}
\caption{The spin flip correlation function. The bubble diagram on the right corresponds to ignoring vertex corrections 
when there are no interactions between the $|0\rangle$ and $|1\rangle$ atoms. 
The thick lines are interacting Green's functions.}
\label{RFFeyn}
\end{figure}

 In absence of interactions, the transition rate exhibits a strong enhancement as the frequency $\omega$ of the RF field is scanned in the vicinity of the free $|0\rangle-|1\rangle$ transition energy. Interactions between the atoms will broaden and shift the resonance peak, and  the RF spectrum therefore provides  information on  interaction effects.
Generally, all three interspecies scattering lengths $a_{\up 0}$, $a_{\up 1}$, and $a_{01}$ may be sizeable. The calculation of ${\mathcal{D}}(\omega)$ is greatly simplified when there are no interactions between the $|0\rangle$
 and the $|1\rangle$ atoms~\footnote{A large $a_{01}$ significantly complicates the calculation of $\mathcal{D}$
and the interpretation of the experiment, since vertex corrections are then important~\cite{Pieri2009}; in the limit of infinitely massive $\ket \up$ atoms with low density, these vertex corrections can be included analytically~\cite{Bruun2010b}.}.
  In this case, vertex corrections can be ignored and the spectral response may be written as
\beq
{\rm Im}{\mathcal{D}}(\omega)=-\frac{1}{2}\int \frac{{\rm d}^3k}{(2\pi)^3}\int \frac{{\rm d}\epsilon}{2\pi}
\left[f(\epsilon)-f(\epsilon+\tilde\omega)\right] A_0(\bk,\epsilon) A_1(\bk,\epsilon+\tilde\omega),
\label{Oneloop}
\eeq
where $f(x)=[\exp(\beta x)+1]^{-1}$ is the Fermi function, $\tilde\omega=\omega+\mu_0-\mu_1$,
and $A_\sigma(\bk,\epsilon)$ is the spectral function of the $\sigma$-atoms
defined as $A_\sigma(\p,\omega)=-2{\rm Im}[G_\sigma(\p,\omega)]$. The spectral function gives the probability  that a $\sigma$ atom with momentum $\p$ has the energy $\omega$. The Feynman diagram corresponding to Eq.\ (\ref{Oneloop}) is shown in Fig.\ \ref{RFFeyn}. 

Formula (\ref{Oneloop}) has the structure of a Fermi Golden rule: the product $A_0A_1$ gives the transition probability between the initial and final states
with momentum ${\mathbf p}$  and energy $\epsilon$ and $\epsilon+\tilde\omega$, while
 the two Fermi functions describe the processes $0\rightarrow 1$ and $1\rightarrow 0$. Since we
have assumed that only very few particles are transferred to state $|1\rangle$, we can ignore the latter process and we obtain
\beq
{\rm Im}{\mathcal{D}}(\omega)=-\frac{1}{2}\int \frac{{\rm d}^3k}{(2\pi)^3}\int \frac{{\rm d}\epsilon}{2\pi}
f(\epsilon) A_0(\bk,\epsilon) A_1(\bk,\epsilon+\tilde\omega),
\eeq
The most clear cut interpretation of the RF probe is when only one interspecies scattering length is significant, either $a_{\up 0}$ or $a_{\up 1}$.
One then has two possible scenarios.

In the first case, termed {\it direct} RF spectroscopy, the magnetic field is tuned close to a $\ket \up-\ket 0$ resonance, and the initial mixture  is strongly
interacting. The RF pulse  probes this system by flipping impurities into a non-interacting state. Then $A_0$ corresponds to the spectral function $A_\downarrow$, while we may take
$A_1(\bk,\omega)=2\pi\delta(\omega-\xi_{k,1})$ since the $\ket 1$  particles are non-interacting.
The spectrum reads therefore:
\beq
{\rm Im}{\mathcal{D}}(\omega)=-\frac{1}{2}\int \frac{{\rm d}^3k}{(2\pi)^3} f(\xi_{k,0}-\omega) A_\downarrow(\bk,\xi_{k,0}-\omega).
\eeq
The advantage of direct RF spectroscopy is that it allows one to probe the ground state of a mixture with arbitrary populations in the two interacting  states $\ket 0$ and $\ket \up$. This scheme has been used successfully to probe pairing phenomena for a balanced mixture in the strongly interacting  BEC-BCS cross-over region~\cite{Chin2004,Shin2007,Stewart2008}, and the attractive polaron energy~\cite{Schirotzek2009}.

In the second case, termed {\it inverse} RF spectroscopy, the magnetic field is instead tuned close to a $\ket \up-\ket 1$ resonance, such that the RF pulse flips
 initially non-interacting minority atoms into a strongly-interacting state. We can then take $A_0(\bk,\omega)=2\pi\delta(\omega-\xi_{k,0})$ which
 gives
\beq
{\rm Im}{\mathcal{D}}(\omega)=-\frac{1}{2}\int \frac{{\rm d}\bk}{(2\pi)^3} f(\xi_{k,0}) A_\downarrow(\bk,\xi_{k,1}+\omega).
\eeq
If there are very few impurities all with the same momentum ${\mathbf k}$, the RF spectrum then becomes simply proportional to the spectral function itself:
\beq
{\rm Im}{\mathcal{D}}(\omega)\propto A_\downarrow(\bk,\xi_{k,1}+\omega).
\eeq
In other words, inverse RF spectroscopy flips particles into a strongly-interacting state thereby directly probing the whole spectral function.
The advantage of inverse RF spectroscopy is that one can study the full excitation spectrum by tuning the RF frequency to flip into all possible states of the system. Inverse spectroscopy turned out to
be the key method to study the excited states of the impurity problem, such as for example the repulsive polaron and the dressed dimer+hole continuum~\cite{Kohstall2012,Koschorreck2012}.


\section{Experimental and theoretical results}
\label{sec:ManyBody}
As we have seen in the previous section, the spectrum of an impurity in a Fermi gas generally presents three branches. At positive energy one finds the repulsive polaron, while
at lower energies  one finds a dressed dimer+hole continuum and the attractive polaron. We review
 in this section the properties of these excitations, studying their dependence as a function of the interaction strength, the value of $k_F R^*$, the impurity mass,  and  the spatial dimension.

\begin{figure}
\includegraphics[width=0.5\linewidth]{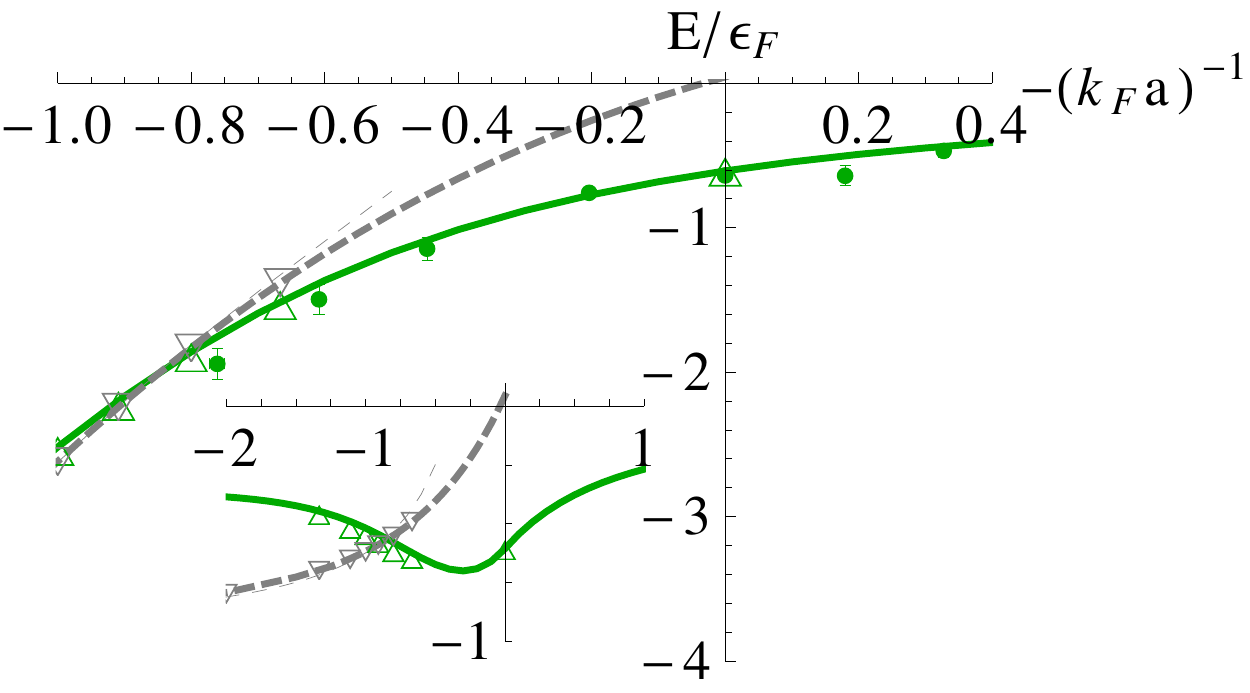}
\caption{Energies of a zero momentum attractive polaron (green) and dressed dimer (gray), as a function of the interaction parameter $-1/k_F a$,
as found respectively by variational/diagrammatic Ans\"atze
(thick lines) and MC calculations (triangles) for a broad resonance ($R^{*}=0$) and equal masses $m_\downarrow=m_\uparrow$.
 The filled symbols are the experimental results of \cite{Schirotzek2009}, corrected to remove final state effects, and the thin line is the mean field result (\ref{molEnergyMF}).
Inset: theoretical data plotted by subtracting the vacuum contribution, i.e., the 2-body binding energy $E_{\rm d}^{\rm vac}$, to highlight the position of the polaron/dressed dimer crossing.
}
\label{fig:polaronAndMoleculeEnergies}
\end{figure}

\subsection{Quasiparticle properties}
The polarons are quasiparticles with an extended spatial structure. A close study of the problem
shows that the wavefunction (\ref{polaronAnsatz}) induces Friedel-like oscillations in the density around the impurity
with period $\pi/k_F$, which are still significant at distances of several $1/k_F$~\cite{Trefzger2013}.
As the envelope of the oscillations decays as $x^{-4}$, the rms radius of the dressing cloud is actually logarithmically divergent. The dressing cloud contains a number of majority particles of order unity, either in excess or missing\ \cite{Massignan2011}.

In the BCS limit, the 1PH ansatz (\ref{polaronAnsatz}) recovers the weak-coupling result $E_-=2\pi a/m_r$. At resonance, the polaron  energy is of order the Fermi energy, see Figs.\ \ref{fig:minoritySpectrum} and \ref{fig:polaronAndMoleculeEnergies}. In the BEC limit, the attractive polaron state  obtained from (\ref{polaronAnsatz})   describes the formation of a dressed dimer with momentum $k_F$ and a hole at the  Fermi surface.
 The energy is therefore   $E_-=E_{\rm d}^{\rm vac}-\epsilon_F+k_F^2/2M=E_{\rm d}^{\rm vac}-\epsilon_Fm_\downarrow/M$ with $E_{\rm d}^{\rm vac}$ the dimer energy    in vacuum (\ref{dimerEnergyInVacuum}). In this regime however, as may be noticed in Figs.\ \ref{fig:minoritySpectrum} and \ref{fig:polaronAndMoleculeEnergies}, the ground state is a dressed dimer with zero momentum.
Its energy can be obtained using the 1PH ansatz given in (\ref{moleculeAnsatz}). In the BEC limit, this ansatz for the dressed dimer recovers the exact result
\beq
E_{\rm d}=E_{\rm d}^{\rm vac}+2\pi a_{ad}n_\up/m_3-\epsilon_F,
\label{molEnergyMF}
\eeq
 where
 $a_{ad}$ is the atom-dimer scattering length obtained from the Skorniakov Ter-Martirosian equation~\cite{Skorniakov1957,Petrov2003}, and $m_3=m_\up(m_\up+m_\down)/(2m_\up+m_\down)$ is the atom-dimer reduced mass. For the equal masses case, one finds $a_{ad}=1.18a$.

\begin{figure}
\includegraphics[width=0.5\linewidth]{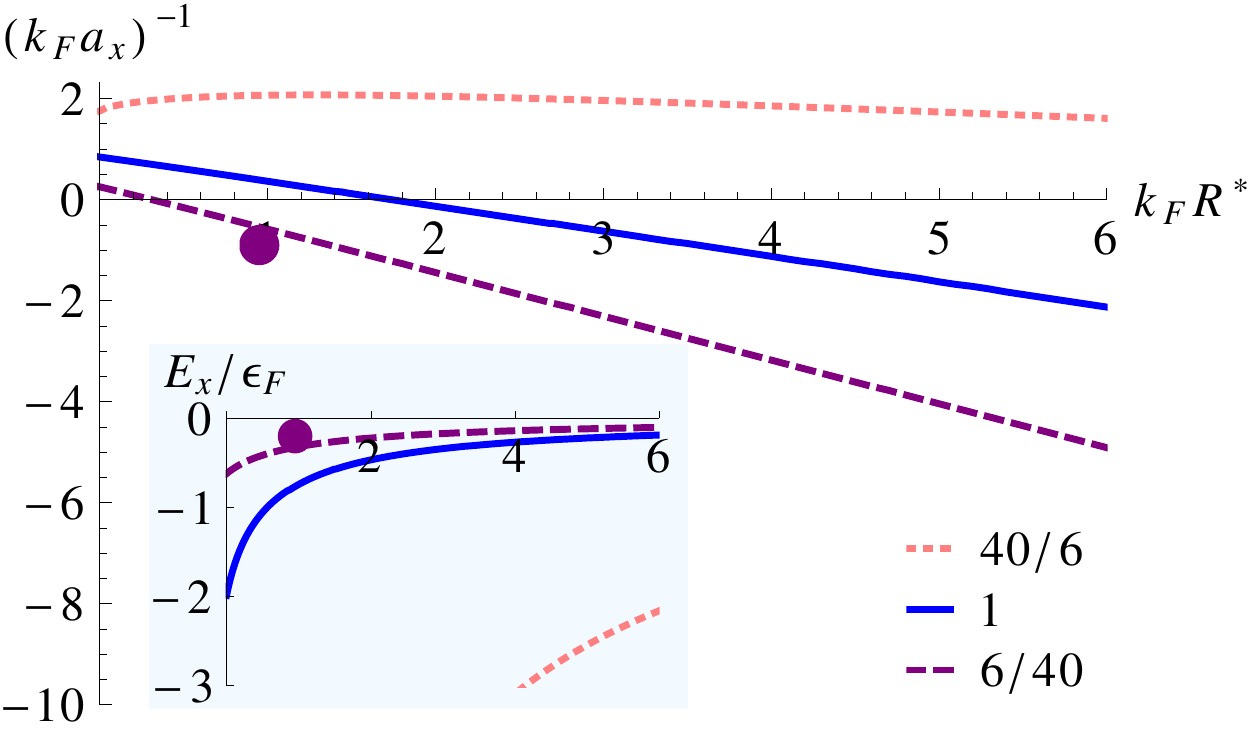}
\caption{Critical interaction strength of the polaron/dressed dimer crossing as a function of the resonance width. From top to bottom, lines are for mass
 $m_\up/m_\down=$40/6, 1, and 6/40. Above the line the ground state is a dressed dimer, below it is a polaron.
 Inset: energy $E_x$ of the excitations at the crossing. The dot marks the interaction strength and energy of the crossing as located in the K-Li mixture of
  Ref.~\cite{Kohstall2012}. Reprinted from \cite{Massignan2012}.
 }
\label{fig:criticalInteractionAndEnergyAtThePolMolCrossing}
\end{figure}

The two branches at negative energies cross in the vicinity of the resonance at a critical coupling strength $1/k_Fa_x$.
We highlight this crossing in Fig.\ \ref{fig:polaronAndMoleculeEnergies} by subtracting the
vacuum dimer energy $E_{\rm d}^{\rm vac}$ from the polaron and dressed dimer energies.
Beyond $1/k_Fa_x$, the attractive polaron energy is inside the dressed dimer+hole continuum, and it becomes energetically favorable for the minority atom to pick out a majority atom and to form a dressed dimer. As a consequence, the ground state quasiparticle changes character across this point, going from a polaron to a dressed dimer.
 The  crossing occurs at $1/k_Fa_x=0.9$ for a broad resonance and equal masses~\cite{Prokofev2008,Mora2009,Punk2009,Combescot2009}, and moves to the BCS side with increasing range
parameter $R^*$~\cite{Massignan2012,Qi2012,Trefzger2012}.
  This is illustrated in Fig.~\ref{fig:criticalInteractionAndEnergyAtThePolMolCrossing} which shows the critical coupling for the polaron/dressed dimer crossing, and the associated energy, as a function of the range $k_FR^*$ for various mass ratios $m_\uparrow/m_\downarrow$.

As we can see from Fig.\ \ref{fig:polaronAndMoleculeEnergies}, the
 1PH approximarion  turns out to be surprisingly accurate for the attractive polaron energy  when compared with the QMC calculations, even in the strongly interacting regime  $k_F|a|\gg 1$. This important discovery of a simple yet quantitatively accurate theory for a strongly interacting many-body problem was only obtained within the last few years with the advent of cold atom gas experiments, since the highly polarised limit is virtually impossible to achieve in  condensed matter systems.
Corrections to the 1PH theory may be evaluated by computing the next order, i.e., the two-particle-hole corrections $\Sigma^{(2)}(p,E)$ in Eq.\ (\ref{SelfEnergy}).
The corrections proved  to be very small with the energy very rapidly converging to the QMC result \cite{Combescot2008}.
By analysing the contribution of diagrams with a higher number of particle-hole excitations, it was shown that the surprising accuracy of the 1PH approximation is most likely due to a fortunate cancellation of higher order diagrams~\cite{Combescot2008,Vlietinck2013}, a phenomenon referred to as {\it sign blessing}. This accuracy should   be compared with the unpolarised case with $N_\uparrow=N_\downarrow$, where the ladder approximation yields  less accurate results and it is much harder to develop accurate theories \cite{BCSBEC2012}.
  Note however that one needs to go beyond the one particle-hole approximation to get even qualitatively correct results for the lifetimes of the quasiparticles, as we shall discuss in Sec. \ref{subsec:decay}.

\begin{figure}
\begin{center}
\includegraphics[width=0.5\linewidth]{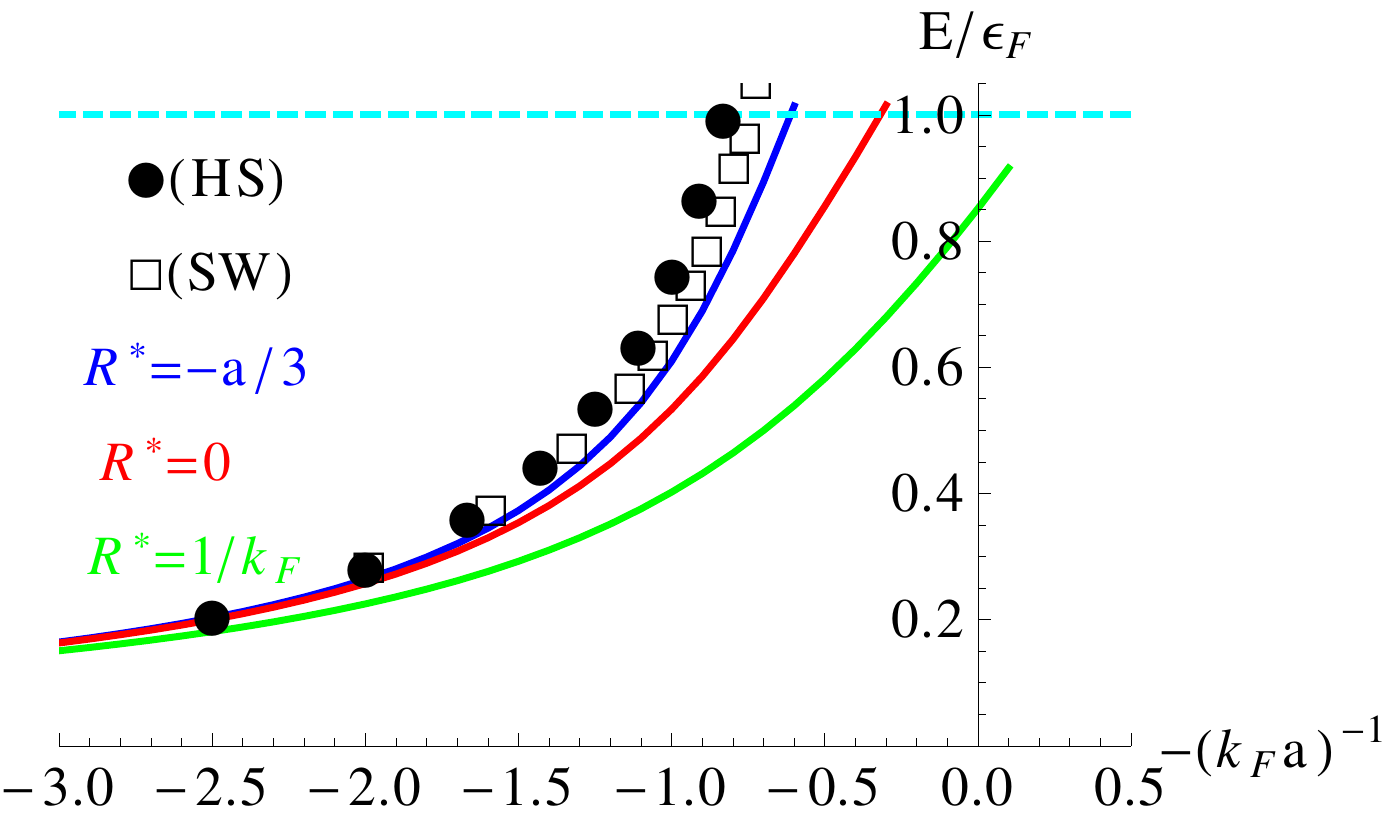}
\end{center}
\caption{Energy of the repulsive polaron as a function of the interaction parameter $1/k_F a$, as found respectively by variational/diagrammatic Ans\"atze for various effective ranges (lines) and MC calculations (symbols); HS and SW stand respectively for hard-sphere and square-well interactions potentials.}
\label{fig:repPolaronEnergy}
\end{figure}
Concerning the repulsive polaron energy, as illustrated in Fig.~\ref{fig:repPolaronEnergy},
there is some discrepancy for strong coupling between the results obtained from the 1PH approximation~\cite{Massignan2011} and the
ones of the Monte-Carlo calculation of~\cite{Pilati2010}. The agreement is rather good when comparing the QMC calculation for hard-spheres (HS) with a diagrammatic calculation with the corresponding range (as the range parameter for HS is $R^{*}=-a/3$), while larger differences appear when comparing diagrammatic results with a QMC calculation based on square-well (SW) potentials (for which $R^{*}\sim0$). The diagrammatic results lie always below the QMC ones.
While the QMC calculation for HS is performed in theground state of ths syetm, and as such can be based on the very reliable fixed-node diffusion Monte-Carlo method (FN-DMC), the repulsive branch of the SW potential is an excited state which may only be studied by variational QMC, which is known to be not as accurate as FN-DMC. A detailed discussion of this point will be given in the following Secs.\ \ref{subsec:IFMrepulsive} and \ref{subsec:IFMattractiveShortRange}.
Of course, the residual difference could also be due to higher order particle-hole processes neglected in the diagrammatic treatment.

\begin{figure}
\includegraphics[width=0.5\linewidth]{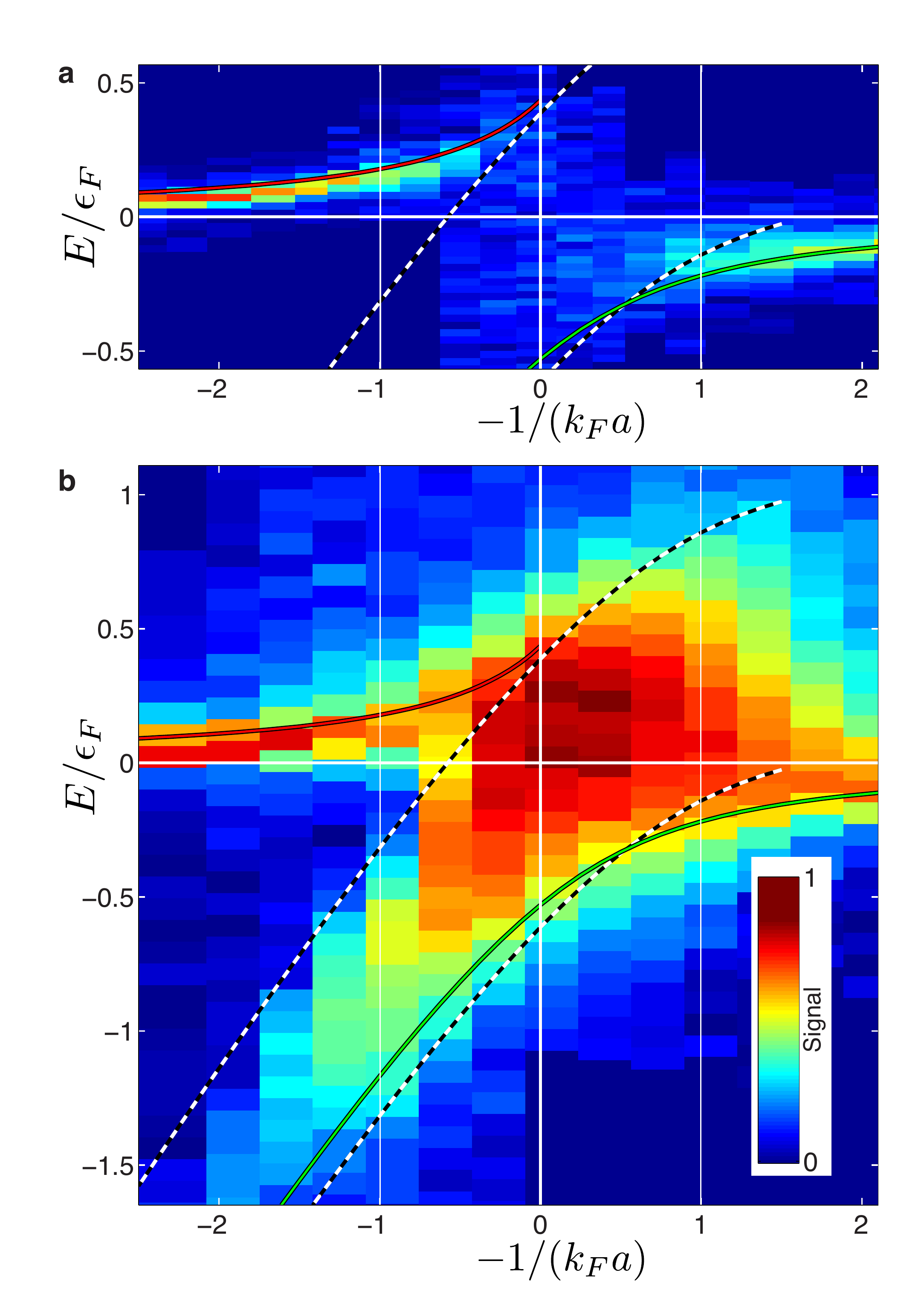}
\caption{Excitation spectrum of $^{40}$K impurities, immersed in a $T\sim0.1T_F$ bath of $^{6}$Li atoms, at a Feshbach resonance with width $k_FR^*\sim 1$, as obtained by inverse RF spectroscopy. The top (bottom) spectrum is obtained using low (high) RF power. The lines are the 1PH theory.
  Reprinted from \cite{Kohstall2012}.
}
\label{fig:spectralFunctionNatureInnsbruck}
\end{figure}

Importantly, the 1PH approximation also agrees with experimental results in the complete spectral range of the polaron problem.
The attractive polaron was first probed in a $^6$Li mixture~\cite{Schirotzek2009} using direct RF spectroscopy. The measured energies are shown in Fig.\ \ref{fig:polaronAndMoleculeEnergies}. This experiment also showed that the polaron-polaron interaction is  weak, in agreement with theoretical calculations~\cite{Mora2010,Yu2010,Giraud2012}.
The whole spectrum including the attractive and repulsive polarons and the
dressed dimer+hole continuum was  investigated by inverse RF spectroscopy in a $^6$Li-$^{40}$K mixture~\cite{Kohstall2012},
and  it is shown in Fig.~\ref{fig:spectralFunctionNatureInnsbruck}. At low RF power (top figure), a significant signal is obtained only from the two polaronic branches, given the small overlap between free impurities and dressed dimers. The dressed dimer+hole continuum becomes on the other hand clearly visible by increasing the intensity of the RF radiation (lower figure). Again, the agreement with theory (lines) is remarkable.

\begin{figure}
\includegraphics[width=0.48\linewidth]{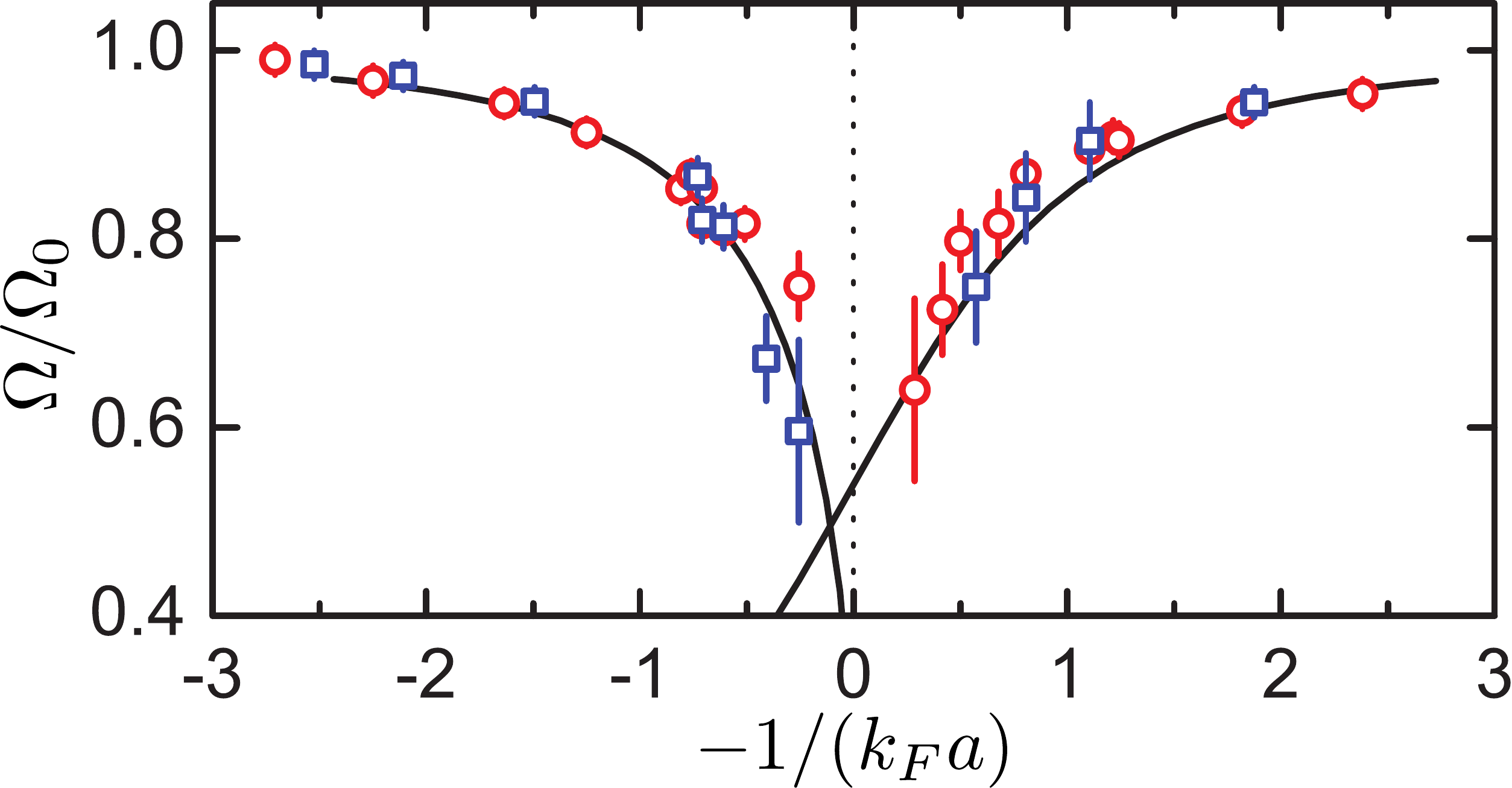}
\caption{Square root of the polaron residues for the $^{40}$K-$^6$Li mixture with $k_FR^*=1$ extracted by the normalized Rabi frequency, $\Omega/\Omega_0$. Lines are the 1PH results for $\sqrt{Z_{\pm}}$, symbols the experimental measurements. Reprinted from \cite{Kohstall2012}.}
\label{fig:polaronResidue}
\end{figure}

 The excellent agreement between the 1PH ansatz, QMC and experiments even holds at the level of the wave function. Figure \ref{fig:polaronResidue} shows that the results for the polaron residues $Z_\pm$ as obtained from the study of Rabi oscillations lie right on top of the 1PH theory predictions \cite{Kohstall2012}. Recent calculations also confirmed that the residue returned by the 1PH ansatz is essentially indistinguishable from accurate QMC results \cite{Vlietinck2013}.
The effective mass of the polarons has been measured by the Paris group in Refs.\ \cite{Nascimbene2009,Navon2010} in three-dimensional system, and by \cite{Koschorreck2012} in two dimensions. At unitarity in three dimensions, experiments and theories agree in locating the effective mass of the attractive polaron in the range $1.1<m^*_-/m_\down<1.2$.

Moving polarons scatter with the surrounding Fermi sea, radiating particle-hole pairs in their neighborhood and decaying to lower momenta. The collisional damping rate of moving polarons was obtained within Fermi liquid theory by \cite{Bruun2008}, and at small momenta it scales as $(k/k_F)^4$ for a single impurity. In the weakly interacting regime, the energy, collisional damping, and effective mass of a moving attractive polaron have been calculated analytically to 2$^{\rm nd}$ order in $k_Fa$ in \cite{Trefzger2013a}. The collisional damping rate of the polaron was also calculated in Ref.~\cite{Baarsma2012}.

Analytical results may also be found in another regime, the one of a very narrow resonance where $k_FR^*\gg 1$. Here, a two-component Fermi mixture is
 weakly-interacting as the coupling between open and closed channels becomes proportional to $(k_FR^*)^{-1/2}$, allowing for a controlled perturbative expansion in this small parameter \cite{Gurarie2007}. For the impurity problem, the critical interaction strength for the transition from a polaron to a dressed dimer may be calculated analytically, yielding~\cite{Trefzger2012}
\beq
\frac 1{k_Fa_x}=-\rho k_FR^*+\frac 2 \pi\left[1-\rho^{-2} + \frac 1 2\left(\rho^{-5/2}-\sqrt{\rho}\right)\log\frac{1+\sqrt \rho}{1-\sqrt\rho}\right]+
{\mathcal O}\left(\frac{1}{k_FR^*}\right)
\label{criticalInteractionAtThePolMolCrossing}
\eeq
with $\rho=m_r/m_\up$. 

In the case of a narrow resonance, one also has a nice physical interpretation for the existence of a 
stable 2-body bound state, i.e.\ the dressed dimer, on the BCS side, even though no such state exists in a vacuum. The reason is that the Fermi sea blocks the decay channels of the dimer. This is partly offset by a positive change in the dressed dimer energy due to Fermi blocking, but the net effect is that the dimer is stabilised on the BCS side by the Fermi sea~\cite{Trefzger2012}.

\subsection{One and two dimensions}
 In two dimensions (2D), the spectrum of an imbalanced two-component mixture of $^{40}$K atoms was measured by  RF
 spectroscopy~\cite{Koschorreck2012}. Even though the
  role of particle-hole fluctuations is larger in 2D due to the constant density
of states, good agreement is still found between experimental results and the 1PH theory~\cite{Zollner2011,Parish2011,Schmidt2012,Levinsen2012a,Koschorreck2012,Ngampruetikorn2012}.
 The intermediate quasi-2D regime, which describes in which manner a 3D gas becomes increasingly
  2D-like with increasing axial  confinement, was analyzed  in \cite{Levinsen2012a}.

Quasiparticles are well defined in 3D and 2D as long as the impurity mass is finite, whereas phase space arguments show that the quasi-particle picture becomes
 ill-defined in 1D for {\it any} mass ratio (see Sec.\ \ref{sec:largeMassImbalance})~\cite{MullerHartmann1971,Kopp1990,Rosch1995,Castella1996}.
Even though the quasi-particle picture breaks down in 1D, a particle-hole expansion still works reasonably well, its results converging rapidly to exact results obtained from the Bethe ansatz~\cite{McGuire1965,McGuire1966,Giraud2009,Leskinen2010,Doggen2013,
Astrakharchik2013}.
 One finds however that two particle-hole excitations play a larger role when compared with higher dimensions as expected.
In this contest, it is interesting to mention the experimental findings recently obtained  by \cite{Wenz2013}: By employing RF spectroscopy, repulsive interactions were investigated in a 1D elongated trap containing one $\downarrow$ impurity in the presence of a few $\uparrow$ atoms, from 1 to 5. 
Quite surprisingly, it turned out that the exact thermodynamic ($N_\up\gg 1$) result of \cite{McGuire1966} for the energy shift caused by the impurity is very quickly recovered 
with $N_\up$ as small as 4 or 5. This remarkable experiment thereby provides an empirical confirmation of the validity of the particle-hole expansion for this problem:  only a very small number of majority atoms is sufficient to dress the impurity and provide polaronic behavior.
 The dynamics of impurities in 1D configurations has been recently theoretically discussed in \cite{Mathy2012,Fukuhara2013,Massel2013}.

\subsection{Large mass imbalance: Anderson orthogonality catastrophe and trimers}
\label{sec:largeMassImbalance}
When the mass of the impurity atom is much larger than the one of the majority atoms, one recovers the  problem of a static impurity in a Fermi sea.
In this limit, there is
 zero overlap between the non-interacting and interacting ground-state, a phenomenon known as the \emph{orthogonality catastrophe}~\cite{Anderson1967,Mahan2000book}.
As a consequence the  quasiparticle residue $Z$, defined in Eq. (\ref{polaronAnsatz}), vanishes, and the spectral function exhibits a power law singularity, signalling the break-down of the quasiparticle picture. The orthogonality catastrophe
is the source of the so-called {\it edge singularities} observed in the spectra
of various materials after the excitation of localized impurities by x-ray photons~\cite{Nozieres1969,Mahan2000book}.
 The reason for the breakdown of the quasiparticle picture  is that the static impurity can excite an infinite number of low-energy particle-hole excitations, causing a complete ``shake-up" of the Fermi sea. As a consequence, a faithful ansatz should contain all n-particle-hole contributions.
  For a finite mass instead, it follows from momentum conservation
that the particle can generate particle-hole excitations with recoil energy only up to $\sim (2k_F)^2/2M$ with $M=m_\uparrow+m_\downarrow$, 
which limits the available phase-space for scattering. The latter argument does not apply in 1D, where there are no quasiparticles for any mass ratio, as mentioned above.
The interesting question of how the polaron properties approach the static impurity physics with increasing ratio $m_\downarrow/m_\uparrow$ was analysed
 in \cite{Trefzger2013}, while \cite{Knap2012,Knap2013} proposed a series of experimental procedures to study new regimes of the static impurity problem with ultracold atoms.

The energy shift of an infinitely massive impurity may be calculated analytically by means of the Fano theorem \cite{Mahan2000book}.
We briefly review the  derivation here. One considers an impurity at the center of a Fermi gas, which for definiteness is enclosed in a sphere of radius $R$.
  For $s$-wave interactions, the general solution of the two-particle problem which vanishes at the boundary $R$ is $\propto \sin[k_j R+\delta(k_j)]$ with $k_j R+\delta(k_j)=j\pi$, where $j\leq n$ is an integer, and $n=k_F R / \pi$. The s-wave scattering phase shift in presence of the impurity is generally given by $k\cot\delta(k)=-a^{-1}-R^*k^2$, while it vanishes in absence of the impurity. The energy shift induced by the impurity is calculated by subtracting the results for the total energy of the system in presence and absence of the impurity. For broad resonances, one finds for the attractive and repulsive polarons the analytic results \cite{Combescot2007,Massignan2011}
\beq
E_{\pm}=\pm\frac{\epsilon_F}{\pi}\left[(1+\pi^2)\left(\pm\frac{\pi}{2}-\arctan y \right)-y\right]
\label{polaronEnergyInfiniteMass}
\eeq
with $y=1/k_Fa$, which gives $E_{\pm}=\pm\epsilon_F/2$ at resonance.

\begin{figure}
\includegraphics[width=0.48\linewidth]{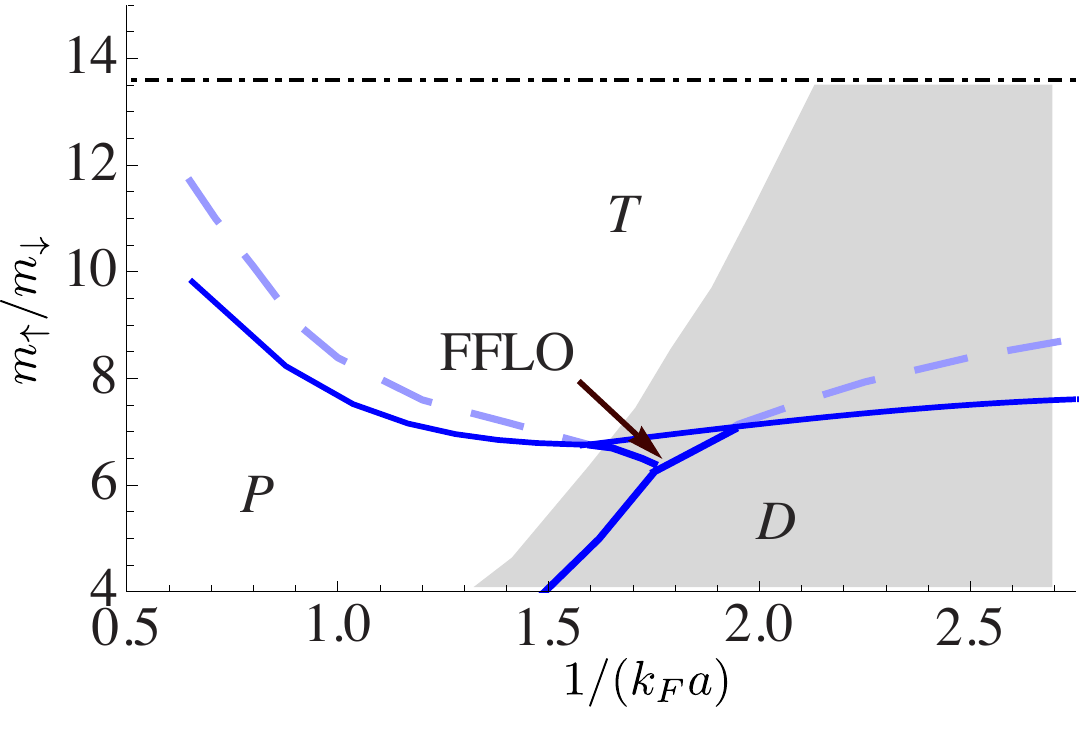}
\includegraphics[width=0.48\linewidth]{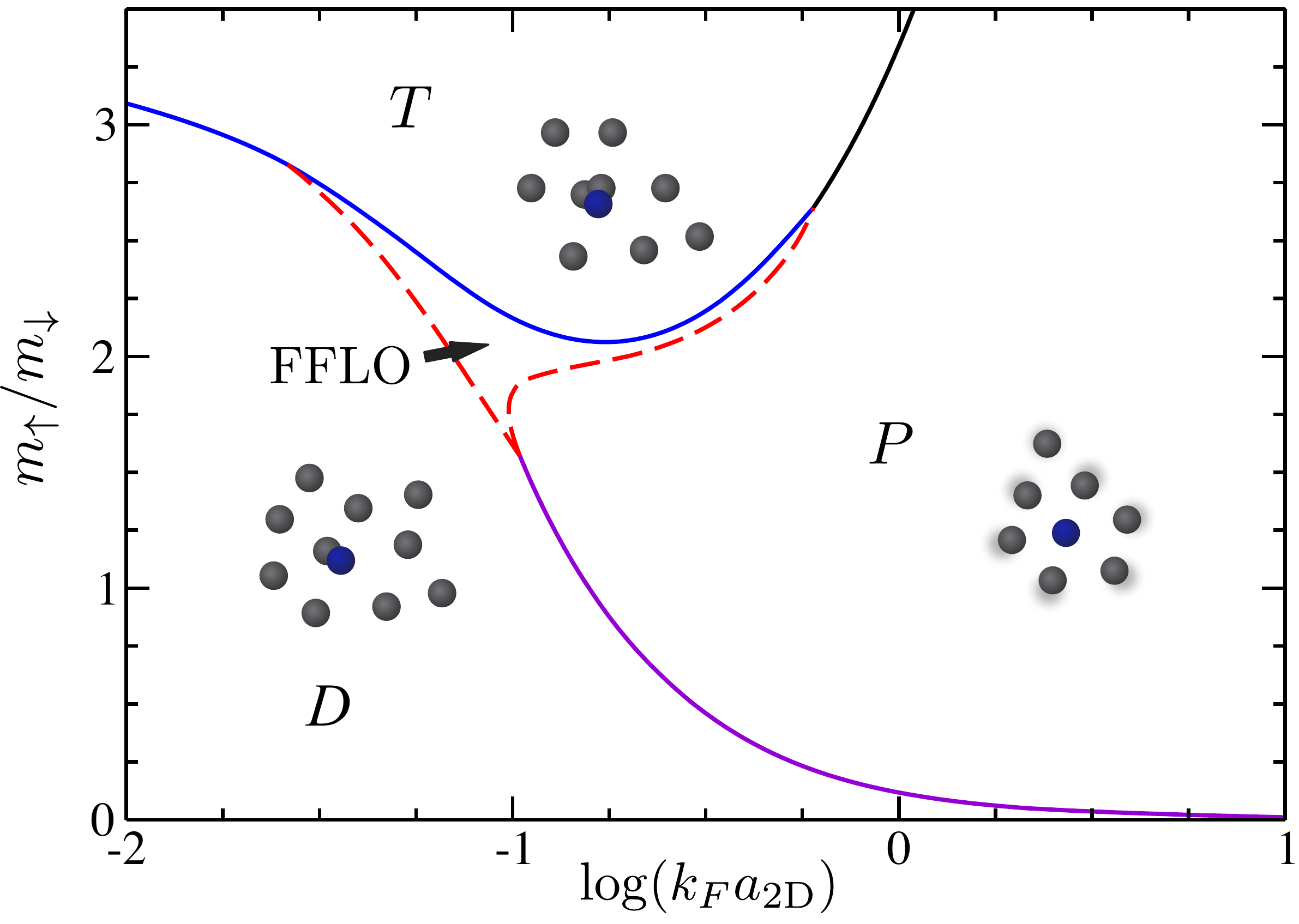}
\caption{Phase diagram of the impurity problem in 3D(left) and 2D(right), as a function of mass ratio and interaction strength. The ground state of a zero-momentum impurity may be a polaron (P), a dressed dimer (D), a dressed trimer (T), or a dressed dimer with non-zero momentum (FFLO). Reprinted from \cite{Mathy2011} and \cite{Parish2013}. \copyright  American Physical Society (APS) 2011, 2013.
}
\label{fig:3Dand2DphaseDiagrams}
\end{figure}

In the opposite limit where the impurities are much lighter than the majority atoms, the formation of a $\down\up\up$ trimer state can be energetically favored.
By generalising the variational wavefunctions (\ref{polaronAnsatz}) and  (\ref{moleculeAnsatz}) to the case of trimers, it was shown that trimers can become the ground state for mass ratios $m_\up/m_\down\ge 6.7$ in 3D \cite{Mathy2011} and $m_\up/m_\down\ge 2.1$ in 2D~\cite{Parish2013}.
The phase diagrams depicting the ground state configuration at a broad resonance in 2D and 3D are shown in Fig.\ \ref{fig:3Dand2DphaseDiagrams} as a function of mass ratio and interaction strength.

\subsection{Quasiparticle decay}
\label{subsec:decay}
While in the toy model of section \ref{subsec:toyModel} 
the excited stable was found to be stable, medium effects present in a many-body system generally couple excited states to lower-lying ones, causing the excitations to acquire a finite lifetime.
 In the context of the impurity problem, the repulsive polaron is unstable towards decay into both the dressed dimer+hole continuum and into the attractive polaron.
Likewise, the attractive polaron can decay into dressed dimer states if $1/k_Fa>1/k_Fa_x$, or viceversa if $1/k_Fa<1/k_Fa_x$. 
 To describe these decay processes, one needs to go beyond the 1PH approximation, since the wave functions (\ref{polaronAnsatz}) and  (\ref{moleculeAnsatz})
 do not couple zero momentum polarons with zero momentum dressed dimers.
  For example, the best the ansatz (\ref{polaronAnsatz}) can do is form a dressed dimer with a majority atom at the Fermi sea as explained above.
   Due to momentum conservation, the coupling between a zero momentum polaron and a zero momentum dressed dimer involves the creation of at least one additional particle-hole excitation in the Fermi sea.
Thus, the leading channels for the decay of an attractive polaron into a dressed dimer, and of a dressed dimer into an attractive polaron, are three-body processes.
The decay of the repulsive polaron into the attractive polaron is also not included in the ladder approximation, as the self-energy (\ref{Selfenergy}) only contains bare minority propagators, while the final state of the decay, the attractive polaron, is itself a quasiparticle.
 A systematic calculation of the polaron decay is  complicated since it involves three-body processes in a many-body medium.

  We outline here a pragmatic and much simpler approach including the most important decay channels, based on the quasiparticle picture with well-defined polarons and dressed dimers.

\subsubsection{Two-body decay}
In a quasiparticle picture, the repulsive polaron can decay into the attractive polaron creating a particle-hole excitation in the Fermi surface to absorb the released energy.
 A simple way to analyse this process is to use the pole expansion of the minority Green's function, Eq.\ (\ref{polaronsPoleExpansion}),
in the vicinity of each polaron energy $E_{\pm}$. This approximation can be used in a ladder type calculation, and the  resulting self-energy is  similar to (\ref{Selfenergy}) with the only difference that the pair propagator in ${\mathcal{T}}$ includes the energy and quasiparticle residue of the  attractive polaron~\cite{Massignan2011}.
  The rate of the   repulsive to
attractive polaron decay can then be extracted as  $\Gamma_{2}=-2Z_{+}\mathrm{Im}[\Sigma(\mathbf{p}=0,E_+)]$, where the factor $2$ comes from the
 fact that we are considering population decay  instead of  wave function amplitude decay. The same method has been
 applied to calculate the decay of the repulsive polaron in 2D~\cite{Ngampruetikorn2012}.

\subsubsection{Three-body decay}
Both the repulsive and the attractive polaron (for $1/k_Fa>1/k_Fa_x$)
can decay to the dressed dimer via a three-body process creating two holes and a particle above the Fermi surface.
Assuming a dilute gas so that the presence of the medium does not affect the scattering events, 
 the three-body decay rate for a single $\downarrow$ atom can be written as~\cite{Petrov2003}
\begin{equation}
\Gamma_{3}=\alpha\frac{\bar \epsilon_\uparrow}{E_{\rm d}^{\rm vac}}n_{\uparrow}^2
\label{PetrovEq}
\end{equation}
where $\alpha$ is a function of the mass ratio $m_\uparrow/m_\downarrow$ and $\bar \epsilon_\uparrow$ is the average
kinetic energy of the $\uparrow$ atoms.  For a broad resonance and $m_\downarrow=m_\uparrow$, one gets $\alpha\simeq 148a^4/m$ and the three-body decay rate
becomes $\Gamma_{3}\simeq0.025(k_Fa)^6\epsilon_F$ per atom.

A systematic three-body calculation in the presence of a medium is very involved. Fortunately,
three-body decay dominates over the two-body one in the BEC limit  $k_Fa< 1$, where a perturbative calculation is reasonable.
We outline this calculation here using Fermi's Golden rule.
The initial state $|I\rangle$ is a zero momentum  polaron described by (\ref{polaronAnsatz}).
The possible final states for the decay consist of two holes on the Fermi surface with momenta
$\ve{q}$ and $\ve{q}'$, a majority particle above the Fermi surface with momentum $\ve{k}$, and a dressed dimer with momentum $\ve{q}+\ve{q}'-\ve{k}$.
 Using Fermi's Golden rule and summing over all possible final states, the decay rate of the polaron via three-body processes is
\begin{equation}
\Gamma_3=\pi\sum_{\ve{k}\ve{q}\ve{q}'}|\mathcal{M}|^2\delta\left(\Delta E+\epsilon_{q,\up}+\epsilon_{q',\up}-\epsilon_{k,\up}
-\frac{(\ve{k}-\ve{q}-\ve{q}')^2}{2M}\right).
\label{3BodyDecay}
\end{equation}
The matrix element between the initial and final state is $\mathcal{M}=(g\phi_{\ve{k}\ve{q}}-g\phi_{\ve{k}\ve{q}'})/\sqrt{\mathcal V}$
and $\Delta E$ is the energy difference between the zero momentum polaron and dressed dimer.
The atom-dimer coupling matrix element $g$ can be obtained from the residue of the scattering matrix $\mathcal{T}$ at the dimer pole, and
 in the BEC limit we can use the vacuum result $g^2=2\pi/m_r^2a^*\sqrt{1+4R^*/a^*}$.
From the variational solution one has
$\phi_{\ve{k}\ve{q}}=\sqrt{Z_P}\mathcal{T}(q,E+\xi_{q\uparrow})/(E-\epsilon_{k\uparrow}
+\epsilon_{q\uparrow}-\xi_{\ve{q}-\ve{k}\downarrow})$.

For the attractive polaron, the three-body decay to the dressed dimer is relevant for  $1/k_Fa>1/k_Fa_x$.
Close to the transition point $1/k_Fa_x$ where $\Delta E\ll \epsilon_F$, it follows from energy and momentum conservation that the particle and hole momenta involved in the
decay form an almost equilateral triangle with vertices at the Fermi surface, such that $\bq+\bq'-\bk\approx 0$. As a consequence, it can be shown that (\ref{3BodyDecay}) gives~\cite{Bruun2010}
\begin{equation}
\Gamma_3\sim k_Fa\left(\frac{|\Delta E|}{\epsilon_F}\right)^{9/2}\epsilon_F.
\label{92scaling}
\end{equation}
A similar calculation shows that the decay rate of the dressed dimers into the attractive polarons  for  $1/k_Fa<1/k_Fa_x$ follows the same $\Delta E^{9/2}$ scaling.
Thus, polarons and dressed dimers are well-defined quasiparticles in the vicinity of the polaron/dressed dimer crossing, as their lifetime diverges much faster than $1/\Delta E$. This 9/2 scaling law has also been obtained numerically in a calculation based on the functional renormalisation group approach~\cite{Schmidt2011}.

Equation (\ref{3BodyDecay}) can also be used to calculate the three-body decay for the repulsive polaron. Using that $E_+-E_{\rm d}\gg \epsilon_F$, the integrals in (\ref{3BodyDecay}) can be simplified and one recovers a form like (\ref{PetrovEq})~\cite{Kohstall2012}.
The function $\alpha$ is however different since it is perturbative and includes many-body effects.
It depends on $a$ and $R^*$, as well as on the polaron quasiparticle residue.
For a broad resonance and equal masses, one gets $\Gamma_{3}\simeq0.015Z_+^3(k_Fa)^6\epsilon_F$ which differs $\simeq40\%$ from the exact vacuum result (\ref{PetrovEq}) in the BEC regime.

\begin{figure}
\includegraphics[width=0.48\linewidth]{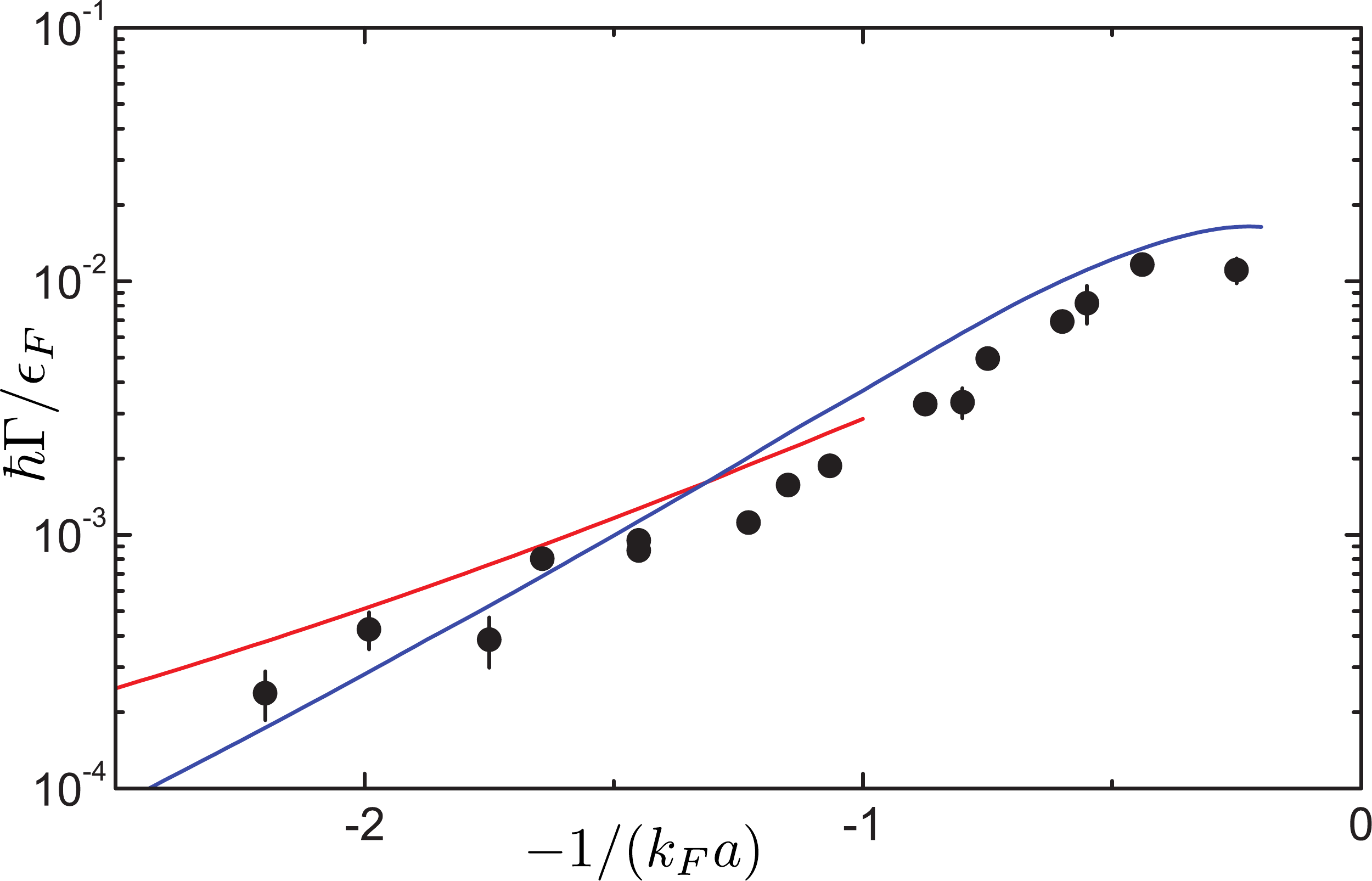}
\caption{Decay rates of the repulsive polaron, as measured employing $^{40}$K impurities in a $^6$Li bath at $k_FR^*=1$. The blue (red) line is the theoretical result for the two-body (three-body) decay discussed in the text. Reprinted from \cite{Kohstall2012}.}
\label{fig:Innsbruckdecay}
\end{figure}

The lifetime of the repulsive branch as a function of the interaction have been measured in the experiment reported in Ref.\ \cite{Kohstall2012}. The
results are shown in Fig.\ \ref{fig:Innsbruckdecay}. The observed decay rates are in good agreement with
the theory outlined above, which notably contains no free parameter. A remarkable feature of the
observed lifetimes is that they are an order of magnitude longer than the ones measured in a balanced $^6$Li mixture at a broad resonance~\cite{Jo2009}.
Indeed, the combination of a large impurity mass and a large range parameter has a very beneficial effect on the lifetime of the repulsive branch. We will discuss this key point in detail in Sec.\ \ref{subsec:StabilityFerro}, in the context of itinerant ferromagnetism.

\subsection{Impurities at p-wave resonances}

Most of the experimental and theoretical studies on atomic quantum gases have been performed at s-wave Feshbach resonances, which are easy to access because these can be very broad in magnetic field; moreover, attractive two-component Fermi gases in the vicinity of a broad s-wave resonance are generally significantly more stable against three-body losses, as compared to the p-wave case. Indeed, weakly-bound dimers at broad s-wave resonances are open-channel dominated, and as such are very large in size and present a small overlap with the deep closed channel ones, which represent the main decay channel. On the other hand, p-wave resonances are intrinsically narrow due to the presence of a short-distance repulsive centrifugal barrier \cite{Landau1977book,Pricoupenko2006}, and as a consequence the weakly-bound p-wave dimers have a large overlap with small and deep closed channel dimers, thereby enhancing three-body decay processes.

\begin{figure}
\includegraphics[width=0.48\linewidth]{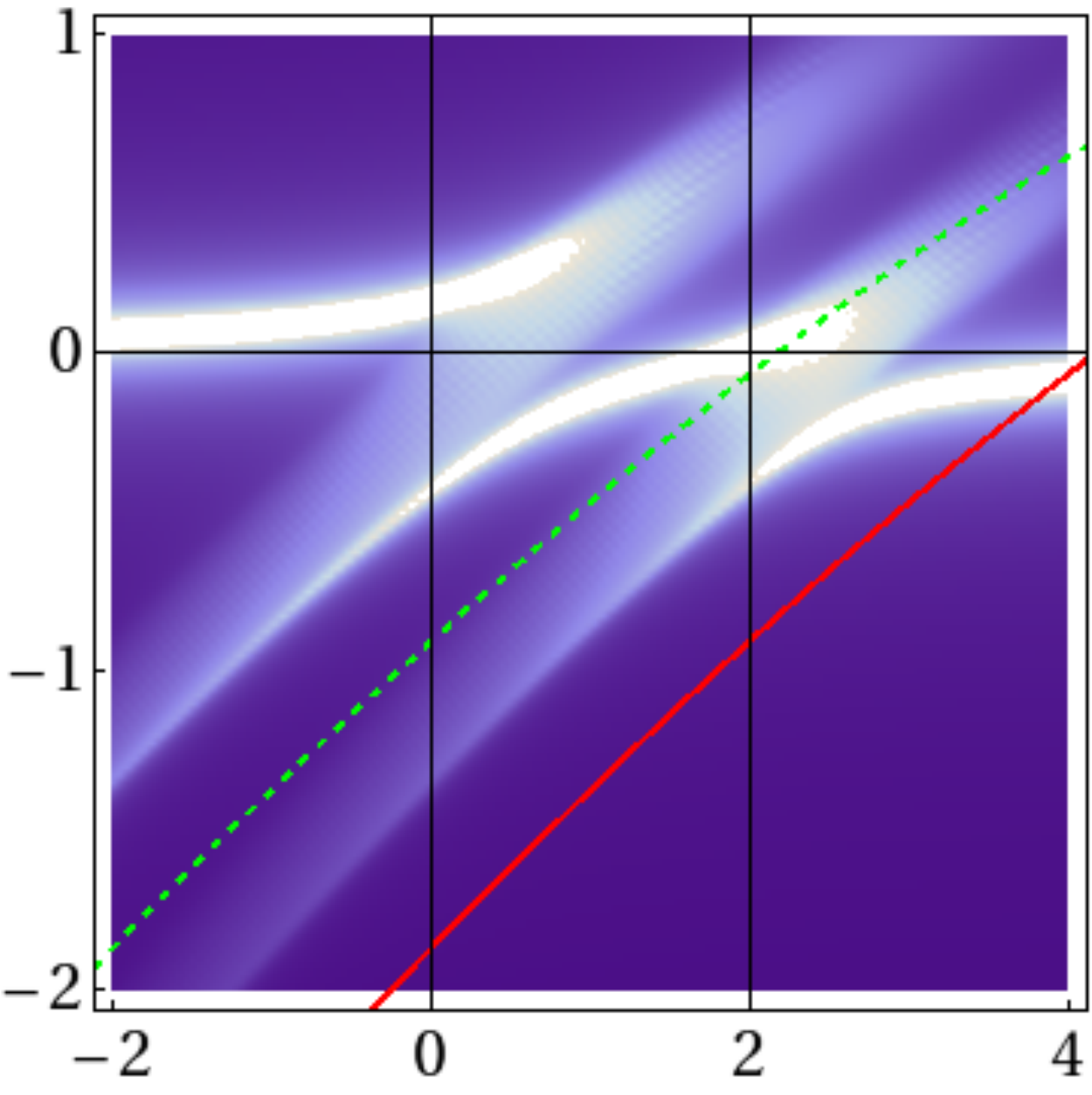}
\caption{Spectral function of the $p$-wave polaron. The horizontal and vertical axes are the interaction parameter $-(k_F^3v_{\pm 1})^{-1}$, with $v_{\pm 1}$ the p-wave scattering volume in the $m_l=\pm1$ state, and the energy $E/\epsilon_F$. The $m_l=\pm 1$ and $m_l=0$ resonances are located respectively at $x=0$ and $x=2$. Thick lines are the dressed molecules with $m_l=\pm 1$ (dashed) and $m_l=0$ (solid).
Results are shown for a p-wave effective range $k_0=-10k_F$, and equal masses of the impurity and the majority particles.
   Reprinted from \cite{Levinsen2012b}. \copyright APS 2012.}
\label{fig:pWavePolaronSpectrum}
\end{figure}

Nonetheless, fermionic gases with p-wave interactions are very interesting, as their phase diagram is predicted to be much richer than the s-wave one, containing polar and chiral phases, and displaying a quantum phase transition between the BEC and BCS regimes \cite{Gurarie2007}. Experimentally, p-wave resonances in either $^6$Li or $^{40}$K gases proved to have very small magnetic widths \cite{Zhang2004,Ticknor2004}, but much wider resonances with magnetic widths up to 10G were later found in a $^6$Li-$^{40}$K mixture \cite{Wille2008}. A peculiar feature of p-wave resonances is the presence of an energy splitting between the dimers with different projections $m_l$ of the relative angular momentum on the magnetic field axis, the dimers with $m_l=\pm 1$ having energy larger than the ones with $m_l=0$ due to dipole-dipole interaction in presence of an external magnetic field \cite{Ticknor2004}. As a consequence, when the energy splitting between the two molecular branches becomes larger than the width of the dressed dimer+hole continuum $\sim\epsilon_F$, the spectrum of an impurity interacting in p-wave with a Fermi sea contains an extra polaron branch and an extra dressed dimer+hole continuum, appearing between the usual attractive and repulsive branches present in the s-wave case \cite{Levinsen2012b}.
 A plot of the polaron spectrum in the latter case is shown in Fig.\ \ref{fig:pWavePolaronSpectrum}. On the contrary, when the energy splitting between the $m_l=\pm1$ and $m_l=0$ dimers is smaller than $\sim\epsilon_F$, the p-wave polaron spectrum closely resembles the s-wave one, displaying a repulsive and an attractive polaron, separated by a single dressed dimer+hole continuum.

At small momentum, the p-wave dispersion may be written as
$E(\p)=E(\0)+p_\parallel^2/2m_{\parallel}^*+p_\perp^2/2m_{\perp}^*$,
where $p_\parallel$ and $p_\perp$ are the projections of $\p$ along
the magnetic field, and on the plane perpendicular to it. At weak-coupling, the effective mass tensor $m^*$ may be computed analytically, yielding $m/m_{\perp, \parallel}^*\approx 1+v_{\perp, \parallel} k_F^3/\pi$ with $v_{\perp, \parallel}$ the p-wave scattering volumes. Interestingly, the effective mass of the polaron may become  smaller than the bare mass $m$. This is possible however only in the regions where the polaron is no longer the ground state, such as on the BEC side of the resonance.

The knowledge of the quasiparticle properties of the p-wave polaron provides access to the equation of state of an imbalanced Fermi gas in its normal phase, as the latter may be accurately described as a mixture of two ideal gases of quasiparticles. In close analogy with the s-wave case discussed later in Eq.\ (\ref{Landau-Pomeranchuk-eqState-sWave}), the equation of state for the p-wave case reads \cite{Levinsen2012b}
\begin{equation}
E=\frac35 \ef N_\up\left[1+\frac {m }
    {(m_\perp^*)^{2/3}(m_\parallel^{*})^{1/3}}\left(\frac{N_\down}{N_\up}\right)^{5/3}\right]+N_\down
E(\0)+\ldots,
\label{Landau-Pomeranchuk-eqState-pWave}
\end{equation}
where $E(\0)$ and $m^{*}$ are the polaron energy and effective mass tensor on the
branch of interest, and $N_{\up,\down}$ the number of spin-$\up,\down$
particles.

\section{Itinerant ferromagnetism in ultracold Fermi gases}
\label{sec:IFM}

In the previous sections, we  described the features of the repulsive polaron consisting of  a single
 $\downarrow$ atom interacting repulsively with a Fermi sea of $\uparrow$ atoms.
We  now consider what happens when the concentration of the impurities is finite in the thermodynamic limit. In particular, we will focus on itinerant ferromagnetism (IFM) in the context of atomic Fermi gases.

Ferromagnetism is a fundamental phenomenon occurring in many solid state  systems, including metals and insulators \cite{Ashcroft1976}. As a consequence of the
 electron-electron interaction, such systems undergo a phase transition at a  critical temperature $T_{c}$:  Upon lowering the temperature below $T_{c}$, the individual magnetic
 moments present in the system, initially thermally disordered, get oriented in certain preferential directions, making ordered patterns. In ferromagnets this leads to a nonvanishing total
 magnetic moment, or "spontaneous magnetization", even in absence of an external field. In antiferromagnets, although there is no net magnetization in the absence of a field, there is a
 far from random spatial distribution of the individual magnetic moments, magnetic interaction favoring in this case antiparallel orientations of neighboring moments.
 In contrast to weak-coupling effects such as  BCS superconductivity,
  the interaction required to drive a ferromagnetic transition is generally \emph{strong}. As a consequence, our understanding of magnetism
  is  characterised by  many open and controversial questions. The problem is best understood in insulators, where the magnetic moments are localized in
  the lattice sites.

  Much more complex is the description of {\it itinerant ferromagnets}, i.e., systems which acquire non-zero magnetization by a spatial re-organization of their mobile magnetic moments. In this category one finds for example compounds such as the transition metals, which owe their magnetic properties to delocalized electrons.
A simple mean field model that captures the main features of itinerant ferromagnets was presented long ago by Stoner~\cite{Stoner1933}.
The basic idea relies on the competition between an effective repulsion between oppositely oriented electron spins, which favors parallel alignment of neighboring magnetic moments, and the Fermi pressure.
 By gradually increasing the repulsive interaction, Stoner's theory predicts that an initially unmagnetized system undergoes a second order quantum phase transition, first to a partially, and then to a fully ferromagnetic phase. The basic idea is analogous to the one suggested by F.\ Bloch\ \cite{Bloch1929}, except that in the Stoner's model screened short-range, rather than pure long-range, Coulomb interactions are considered \cite{Ashcroft1976}.
The critical behavior predicted by Stoner was successively confirmed
using renormalization group methods~\cite{Hertz1976}. Later on, it was realized that particle-hole excitations couple to the magnetic order driving the transition to be first order at low temperature~\cite{Belitz1999,Belitz2005}, while it remains second order at higher temperatures.

Examples of both partially and fully polarized ferromagnetic metals can be found in nature:
 Transition metals such as iron, cobalt and nickel are partially polarized, the magnetic moment per atom in the system being a non-integer number of Bohr units,
 whereas compounds such as CrO$_2$ and EuB$_6$  are completely polarized~\cite{Wijn1991}. For what concerns the character of the magnetic phase transition, experiments performed on sufficiently clean materials are consistent with the transition being first order at low temperature~\cite{Belitz2012}. More generally, the current understanding is that \emph{if}  there is a ferromagnetic transition in a clean system in dimensions higher than one,
\emph{and if} this transition occurs at a sufficiently low temperature, \emph{then} the transition is first order.

Despite great theoretical and experimental progress over the last decades, a detailed comparison of experimental observations in solid state systems with microscopic theories still remains a complicated affair, due to the inevitable presence of disorder and intricate band structures in real materials.
As a matter of fact, it is even still debated whether a homogenous electron system, such as the one considered in Stoner's model, becomes ferromagnetic at all.

The availability of cold atomic gases with strong repulsive interactions has the potential to make significant progress on this fundamental problem
 for several reasons~\cite{Duine2005,Zwerger2009}.
 First, the short-ranged atom-atom interaction is relatively simple to describe accurately when compared to   electron interactions in a solid, and its
strength can be tuned at will, as we already discussed. Second, the dimensionality, temperature and amount of disorder can be easily controlled in experiments. Third, the possibility to
realize mixtures with arbitrarily  (pseudo-)spin populations allows
one to examine ferromagnetism in the strongly polarized regime, where the quantitatively accurate theory in terms of polarons is available.
Finally, a number of observables, such as particle density and momentum distribution, kinetic and interaction energy, are easily accessible by means of absorption and phase-contrast imaging, RF and Bragg spectroscopy~\cite{Chin2004,Shin2007,Stewart2008,Veeravalli2008}.

\begin{figure}
\begin{center}\includegraphics[width=0.8\linewidth]{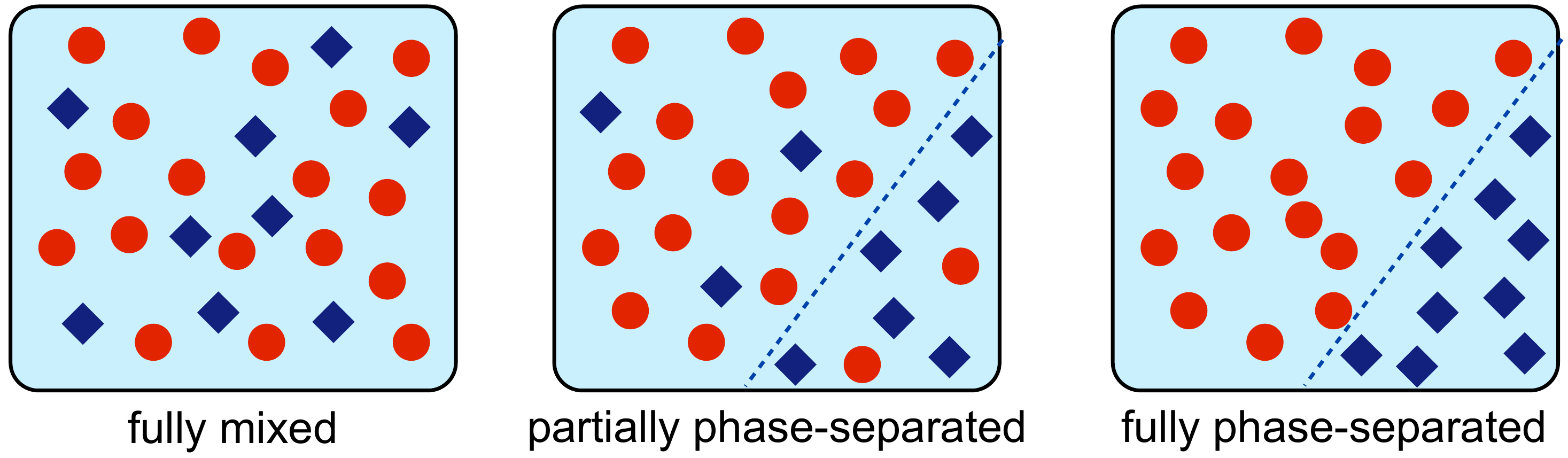}\end{center}
\caption{Spatial configurations of a two-component atomic quantum mixture.}
\label{fig:sketchIFM}
\end{figure}

Two  features of ultracold gases should however be kept in mind when they are employed to study itinerant ferromagnetism. First, in contrast with the case of electrons in solids where only the total electron population is fixed, in ultracold gases spin-changing collisions are rare, and the two pseudospin populations $N_{\uparrow}$ and $N_{\downarrow}$ are generally fixed separately. As a result, the total ``magnetization" $N_\uparrow-N_\downarrow$
 is constrained by the initial number of particles $N_{\uparrow}$ and $N_{\downarrow}$ at which the gas is prepared. In atomic gases, ferromagnetism therefore shows up as the formation  of spatial domains containing a density imbalance $n_{\uparrow}\neq n_{\downarrow}$, and ``saturated ferromagnetism" translates into the creation of spatially
separated domains containing only $\uparrow$ or $\downarrow$ particles. A sketch of these configurations is shown in Fig.\ \ref{fig:sketchIFM}
\footnote{Various authors have also considered more exotic magnetic configurations, such as spin coherent ferromagnetic states \cite{Duine2005}, hedgehog configurations \cite{LeBlanc2009}, and textured phases  \cite{Conduit2009}}.

Second, as we have discussed in section \ref{subsec:decay}
the tunability of the repulsive interaction in ultracold gases comes at a price, since the upper branch of these systems is unstable towards decay.
Consequently, the observation of ferromagnetic phenomena with atomic gases unavoidably competes
with the tendency of the system to relax into the lower-lying ground state.

This Section is organised as follows. We start in Sec.~\ref{sec:perturbation} by looking at the available perturbative treatments for IFM. As these indicate that IFM  is a strong-coupling 
phenomenon, we then switch to consider more elaborate methods. In the strongly-interacting regime, the short-range nature of the interaction potential plays an important role. In Sec.~\ref{subsec:IFMrepulsive} we look at the case of purely repulsive potentials, which can be addressed with very high accuracy by quantum Monte-Carlo methods (QMC). In Sec.~\ref{subsec:IFMattractiveShortRange} we focus  on the experimentally relevant case of repulsive interactions induced by attractive short-range potentials. We first derive a phase diagram in terms of the polaron theory discussed in the earlier sections, and compare its results to QMC ones. We proceed in Sec.~\ref{subsec:StabilityFerro} by showing how the polaron theory offers important information on the lifetime of the upper branch section, and in Sec.~\ref{subsec:IFMexperiments} we discuss the experimental results available in the literature.

\subsection{Perturbation theory}\label{sec:perturbation}
Let us first review a perturbative description of itinerant ferromagnetism.
To first order, this corresponds to the Stoner model.
We consider a homogeneous two-component fermion mixture consisting of $N_\sigma$ atoms with mass $m_\sigma$ with $\sigma=\uparrow,\downarrow$ confined in a total volume $V$.
For simplicity, we  restrict ourselves to the case $m_\up=m_\down=m$ in this subsection, and we consider the broad resonance case ($k_F R^* \ll 1$) where the
 interaction is completely characterized by the scattering length $a$.
To make a  connection with our earlier discussion of the extremely imbalanced case $N_\up+1_\down$, it is useful  to define the Fermi energy $\epsilon_F= k_F^2/2m_\up$ in terms of the Fermi momentum $k_F$ of a {\it single-component} Fermi gas with total density $n=(N_\uparrow+N_\downarrow)/V=N/V$, i.e., $k_F^3=6\pi^2n$.
To compare with existing literature which mostly focuses on balanced mixtures, we also define the Fermi momentum of a balanced two-component
 Fermi gas $k_{F\sigma}^3=3\pi^2n$, yielding the relation $k_{F\sigma}=k_F/2^{1/3}$. Finally, we introduce the polarization $P=(N_{\uparrow}-N_{\downarrow})/(N_{\uparrow}+N_{\downarrow})$.
 
In the absence of interactions, the particles  form a  mixture of two ideal Fermi gases, and the kinetic energy per particle at zero temperature reads
\begin{equation}\label{E0tot}
\epsilon_{\rm kin}= \frac{3}{5} \epsilon_F\left[(1-y)^{5/3}+y^{5/3}\right].
\end{equation}
with  $y=N_\down/N$. In perturbation theory, the interaction energy per particle is calculated in powers of $k_Fa$.
To first order, we recover the Stoner mean field  expression
\begin{equation}\label{StonerInta}
 \epsilon_{\rm int}= \frac{4 \pi a}{m} \frac{n_\up n_\down}{n}=\frac{4}{3 \pi}k_{F}a\epsilon_Fy(1-y).
\end{equation}
Minimizing the total energy in the mixed phase $\epsilon_{\rm kin}+\epsilon_{\rm int}$ with respect to $y$ gives three cases:
(i) for $k_{F\sigma}a=2^{-1/3}k_Fa<\pi/2$, the minimum energy is for $y=1/2$ and the state is non-magnetic; (ii) for
$\pi/2<k_{F\sigma}a< 3\pi/2^{7/3}$, the energy is minimized for $0<y<1/2$
and the system is in a partially  ferromagnetic state; (iii) for $k_{F\sigma}a>3\pi/2^{7/3}$, the energy has a minimum at $y=0$
and the system is in the saturated ferromagnetic state (note that the energy is symmetric under the exchange $y\leftrightarrow 1-y$).
These inequalities give rise to the phase boundary marked as the black line ``1$^{\rm st}$" in Fig.~\ref{fig:phaseDiagram_HS}.
The condition $k_{F\sigma}a>\pi/2$ for ferromagnetism can be expressed as
\beq
g(k_{F\sigma})U>1,
\eeq
 where $g(\epsilon_F)=mk_{F\sigma}/2\pi^2$ is the single component density of states
at the Fermi surface and $U=4\pi a/m$ is the strength of the  zero-range interaction potential. Written in this form, one can easily include changes in DOS due to band-structure effects.

\begin{figure}
\begin{center}
\includegraphics[width=0.5 \columnwidth]{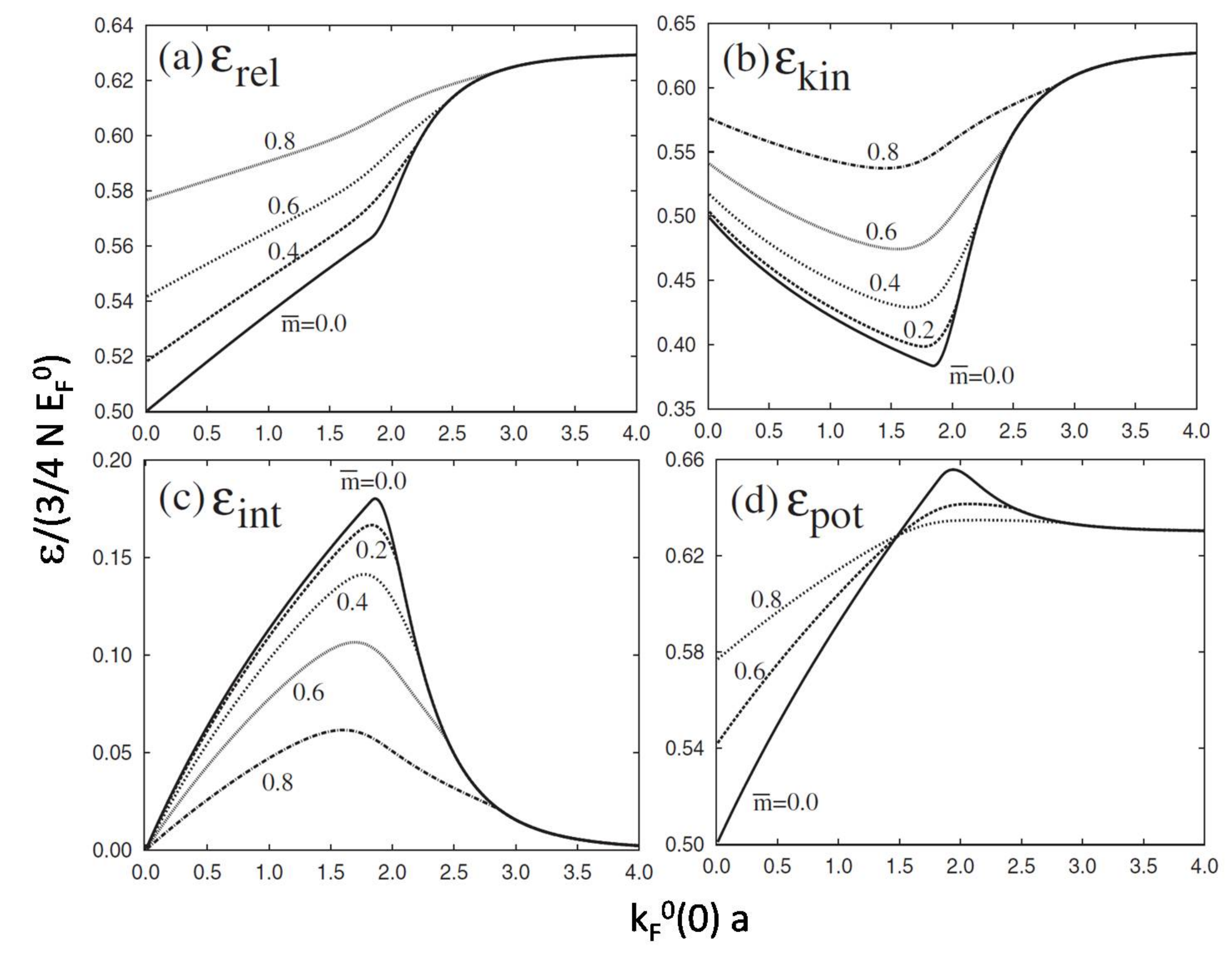}
\end{center}
\caption{Energetics of a two-fermion mixture in a 3D trap at zero temperature, as given by mean field theory. The release \textbf{(a)}, kinetic \textbf{(b)}, interaction \textbf{(c)} and potential \textbf{(d)} energy of a trapped system as a function of the interaction strength,  using the local density approximation. $E_F^0$ is the Fermi energy of a non-interacting Fermi gas of $N/2$ particles, and $k_F^0(0)$ denotes the Fermi wavevector $k_{F\sigma}$ at the trap center of the non-interacting system. $\bar{m}=(N_{\uparrow}-N_{\downarrow})/(N_{\uparrow}+N_{\downarrow})$ denotes the total magnetization. Reprinted from \cite{LeBlanc2009}. \copyright APS 2009.}
\label{fig:MFtrap}
\end{figure}

Stoner's mean field theory has been applied to atomic gases in several early papers~\cite{Houbiers1997,Salasnich2000,Sogo2002,LeBlanc2009}.
Trapped systems can be studied by employing the local density approximation to the Stoner energy functional, and various observables such as the kinetic, potential, interaction and release energies can be evaluated by minimizing the energy for fixed particle numbers $N_\sigma$ (see Fig.\ \ref{fig:MFtrap}). Qualitatively, the behavior of these
quantities is easy to understand. For small repulsion, the mixture is paramagnetic  and the repulsive interaction leads to an
expansion of the two density distributions. This causes an increase of the potential energy of the system, and a decrease of its kinetic energy. At a critical interaction
strength, a polarized region develops at the trap center, yielding a sharp decrease in the interaction energy. Parallel to this, the emergence of phase separation leads to an overall
squeezing of the system, which increases the kinetic energy.

 The Stoner model  provides a qualitative model for itinerant ferromagnetism.  On the other hand, it predicts a transition to a ferromagnetic state at a coupling strength
 so large that it lies outside the range of applicability of mean field theory itself, which requires on general grounds $k_Fa\ll 1$.
Not surprisingly then, the Stoner model fails in quantitatively reproducing the experimental observations on solid state systems. For instance, the Curie temperature $T_c$ calculated by Stoner theory is generally too high (up to a factor $\sim5$) when compared to experiments, and the Curie-Weiss susceptibility above $T_c$ cannot be recovered by this model, except in the high temperature limit where $T>T_F$.
Moreover, the Stoner model  fails  at the qualitative level, since it predicts a second order quantum phase transition  at $T=0$.

A calculation to second order in $k_Fa$ for a homogeneous system has been presented in Ref.\ \cite{Duine2005}. There, it was shown that going to second order in the total energy has two effects. First, it strongly reduces the critical interaction strength for ferromagnetism
(see the two green markers ``PT$_2$" in Fig.~\ref{fig:phaseDiagram_HS}).
 Second, particle-hole excitations drive the transition to become first order for low $T$. This is in agreement with general arguments, which show that the phase transition becomes first order whenever a gapless mode is coupled to the magnetization~\cite{Belitz2005}. 

\subsection{Strong interactions from purely repulsive potentials}
\label{subsec:IFMrepulsive}
\begin{figure}
\begin{center}
\includegraphics[width=0.5 \columnwidth, clip]{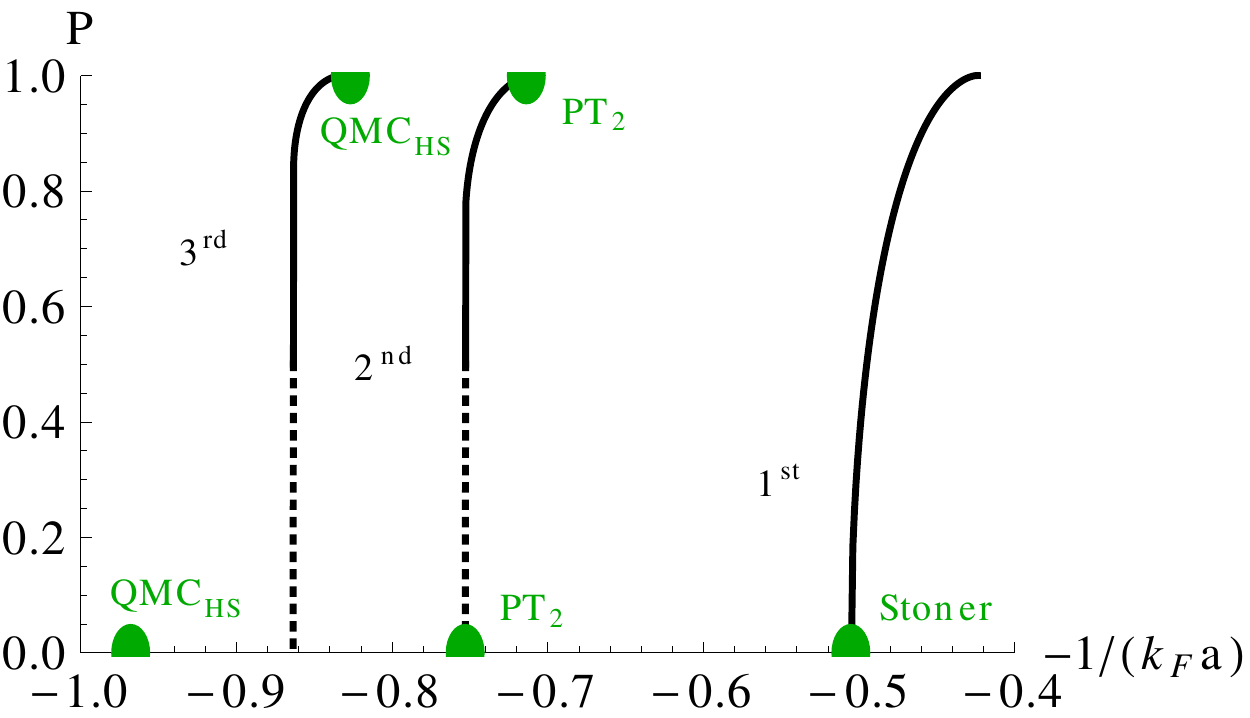}
\end{center}
\caption{Phase diagram of a repulsive Fermi mixture at zero temperature for hard-sphere interactions. The gas is mixed above the lines, and phase separated below. The lines are phase boundaries obtained from the perturbative expression (\ref{BishopEnergy}) retaining terms up to $1^{\rm st}$, $2^{\rm nd}$, and $3^{\rm rd}$ order. The lines are shown as dotted for $P<0.5$, where we do not expect Eq.\ (\ref{BishopEnergy}) to be accurate as impurity-impurity interactions become important. The green markers indicate the IFM transition for $P=0$ and $P=1$ as found by Quantum Monte-Carlo calculations~\cite{Pilati2010}, 2$^{\rm nd}$ order perturbation theory (PT$_2$)~\cite{Duine2005}, and Stoner theory~\cite{Stoner1933}.}
\label{fig:phaseDiagram_HS}
\end{figure}

The reliability of perturbative calculations may be examined by testing the convergence of the result (here, the critical interaction strength) as a function of the
perturbative order. Equation (\ref{BishopEnergy}) provides a perturbative expression in $k_Fa$ for the energy of
 a single impurity interacting via a hard core repulsive potential with  a $T=0$ Fermi sea with density $n$.
Since  we work at fixed density $n=n_\up+n_\down$,  the  density of the Fermi sea is reduced to $n_\up=(1-y)n$ in presence of a finite density
$n_\downarrow=yn$ of impurities,
and the interaction parameter therefore changes to $k_Fa(1-y)^{1/3}$. As long as the density of impurities remains small ($y\ll 1$),
we expect Eq.~(\ref{BishopEnergy}) to be accurate and the interaction energy per particle may  be written as
\beq
\epsilon_{\rm int}=y(1-y)^{2/3}E_+[k_Fa(1-y)^{1/3}].
\label{eInt}
\eeq
 We write $E_+[k_Fa]$ to highlight the  dependence of $E_+$ on the density through  the interaction strength $k_Fa$.
The prefactor $(1-y)^{2/3}$ comes from a reduction of the Fermi energy with $y$.
To first order in the interaction, Eq.\ (\ref{eInt}) reduces to Eq.\ (\ref{StonerInta}). 
We plot in Fig.\ \ref{fig:phaseDiagram_HS} the $T=0$ phase diagram obtained by minimizing the total energy of the mixed phase $\epsilon_{\rm kin}+\epsilon_{\rm int}$ to $1^{\rm st}$, 
$2^{\rm nd}$ and $3^{\rm rd}$ order using Eqs.~(\ref{BishopEnergy}), (\ref{E0tot}), and (\ref{eInt}).
Somewhat surprisingly, we note that when the $2^{\rm nd}$ order perturbative result based on Bishop's formula Eq.~(\ref{BishopEnergy}),
which is derived for a single impurity and therefore strictly valid only in the $P\rightarrow 1$ limit, is extrapolated to $P=0$, it
agrees essentially perfectly with the result found in Ref.~\cite{Duine2005}, which is valid for  all polarizations.
In any case, the significant changes in the diagram as the order of the perturbative expansion is increased highlight the fact that perturbation theory is at best 
 slowly converging for $P=0$ at this strong coupling.

This fact triggered several Quantum Monte Carlo (QMC) studies of the $T=0$ phase diagram of a repulsive Fermi gas.
Within QMC, repulsive pairwise interactions can be modeled by employing either purely repulsive interaction potentials, such as hard sphere (HS) and soft sphere (SS), or
 attractive potentials supporting a bound state, such as square well (SW) and P\"oschl-Teller (PT) potentials.
 Provided the range of the potential is small, the long range physics
  is essentially independent of the microscopic potential employed whereas the short range physics of course varies significantly.
It is important to note that the attractive potentials, such as SW and PT, admit a well-defined zero-range limit with a non-zero scattering length $a$.
In contrast, purely repulsive potentials such as the HS and SS ones are always characterized by a sizeable (negative) value of the range parameter $R^*$, of the order of $a$. For the purely repulsive potentials, the zero range limit corresponds to the trivial non-interacting case. We focus here on the QMC for purely repulsive HS potentials, while we will consider attractive SW potentials in the next Section.

\begin{figure}
\begin{center}
\includegraphics[width=0.5 \columnwidth]{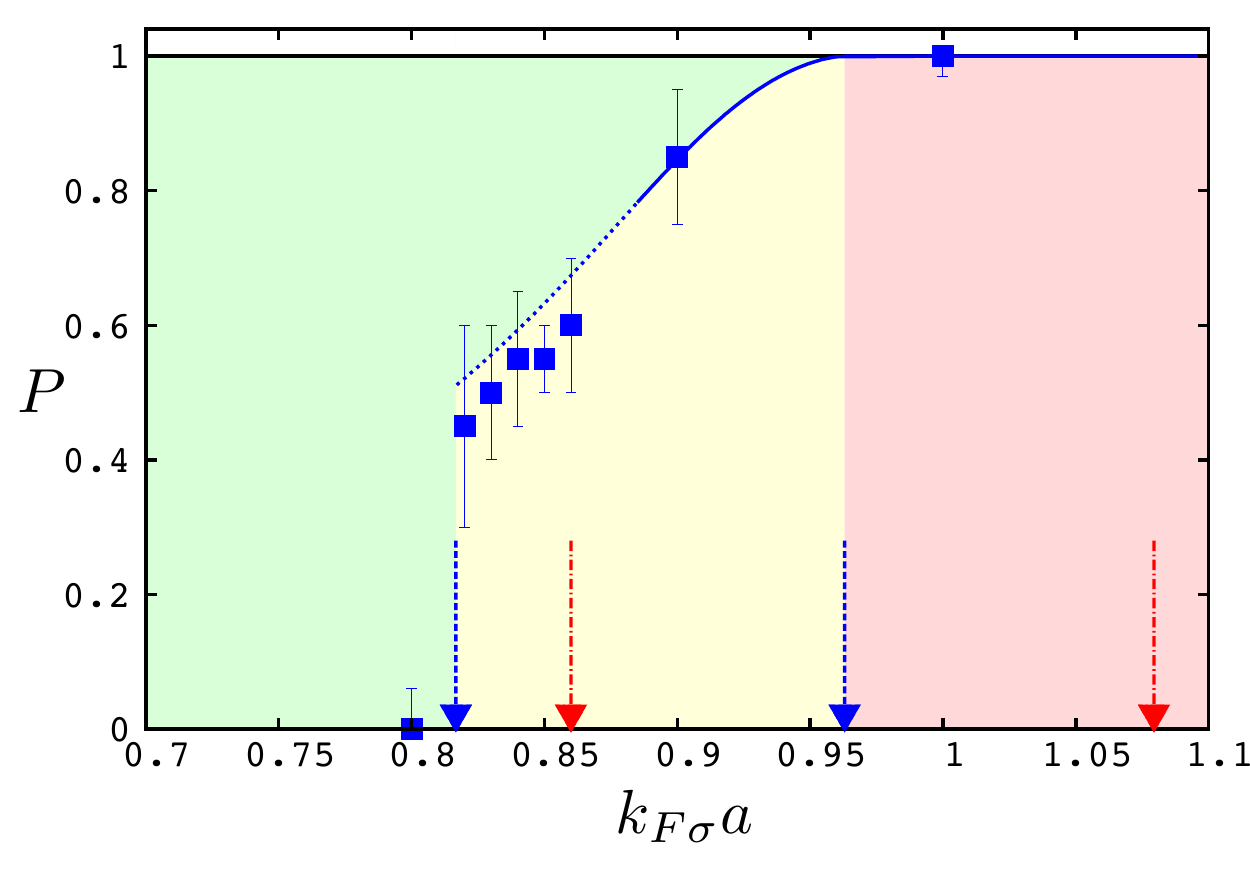}
\end{center}
\caption{
QMC phase diagram of a repulsively interacting fermion mixture at  $T=0$. Three regions are present in the interaction-polarization plane: a homogeneous phase for weak interactions (green), a phase-separated state composed of partially-polarized domains (yellow), and a phase-separated state with fully-polarized domains for strong interactions (pink). The blue symbols correspond to the configuration of minimal energy for the HS potential, and the line is the corresponding phase boundary determined from the equilibrium condition for pressure and chemical potentials. Blue (red) arrows mark the critical values where the magnetic susceptibility diverges and where the fully polarized state becomes stable for the HS (SW) potential. Reprinted from \cite{Pilati2010}. \copyright APS 2010.
}
\label{fig:PilatiPD}
\end{figure}

Particles interacting by means of purely repulsive HS potentials can be studied using fixed-node diffusion Monte Carlo (FN-DMC), which is known to be an essentially exact method to investigate ground state properties~\cite{Pilati2010,Chang2011}.
Using the Monte-Carlo results for the energy as a function of $y=N_{\downarrow}/N$, the $T=0$ phase diagram can be obtained from the Maxwell construction, see Fig. \ref{fig:PilatiPD}. The QMC results for HS potentials at zero and full polarization are also shown as green markers ``QMC$_{\rm HS}$" in Fig.~\ref{fig:phaseDiagram_HS}\footnote{Note the different x-axes in the two figures, where the interaction strength is measured respectively in terms of $k_Fa$, and of $k_{F\sigma}a=k_Fa/2^{1/3}$.}.
We see that the perturbative calculation of Ref.~\cite{Duine2005} approaches the exact result as given by the QMC result, but is still far from having converged. One may also see that the perturbative calculation based on Bishop's expression at third order has instead substantially converged to the exact result at $P=1$. This is consistent with the general conclusion that the
strongly polarized regime allows for a more accurate theoretical description than the balanced case. 

Importantly, Monte Carlo calculations \cite{Pilati2010} carried at arbitrary polarization showed that the equation of state of the system in the mixed phase, for $P\gtrsim 0.5$, is accurately described in terms of a weakly interacting gas of repulsive polarons.
Indeed, in the spirit of the so-called Landau-Pomeranchuk Hamiltonian,  the equation of state for the case $N_\up\gg N_\down$ may be written as
 \begin{equation}
E=\frac35 \ef N_\up\left[1+\frac {m}
    {m_+^*}\left(\frac{N_\down}{N_\up}\right)^{5/3}\right]+N_\down
E_++\ldots,
\label{Landau-Pomeranchuk-eqState-sWave}
\end{equation}
where $E_+$ and $m_+^*$ are the polaron energy and effective mass on the repulsive branch, and $N_{\up,\down}$ the number of spin-$\up,\down$ particles.
 A similar equation in terms of $E_-$ and $m_-^*$ is valid for the gas on the lower branch in the normal phase \cite{Lobo2006}, or in the p-wave case, as we have seen in Eq.\ (\ref{Landau-Pomeranchuk-eqState-pWave}).
The good agreement with the independent QMC calculation at finite density of $\down$ atoms shows once more how the knowledge of the polaron quasiparticle properties provides important insights in the equation of state of a Fermi mixture, even at non-zero population imbalance.

\subsection{Strong repulsion induced by attractive short range potentials}\label{subsec:IFMattractiveShortRange}
The important finding discussed in Sec.~\ref{subsec:IFMrepulsive}, that the mixed phase can be accurately described as a gas of polarons  mixed with an ideal gas of $\uparrow$ atoms when $P\gtrsim 0.5$ (or $y<1/4$), allows one to develop a quantitatively reliable theory for itinerant ferromagnetism in the polarized regime.
This regime has been scarcely investigated before, since strong polarization is virtually impossible to realize with electron systems.

The non-zero temperature version of Eq.\ (\ref{Landau-Pomeranchuk-eqState-sWave}) for the free energy per particle in the mixed phase reads~\cite{Massignan2013}
\beq
\fcal_{\rm mix}(y)=(1-y)\fcal_{\up}(n_\up,T)+y\fcal_{\down}(n_\down,T)+\epsilon_{\rm int}
\label{energyPPmix}
\eeq
where $\fcal_\sigma(n_\sigma,T)=\mu_\sigma-k_BT {\rm Li}_{5/2}(-z_\sigma)/{\rm Li}_{3/2}(-z_\sigma)$ is the free energy per particle of an ideal gas
of $\sigma$-atoms at temperature $T$, and ${\rm Li}_x(z)$ is the polylogarithm function of order $x$.
 The fugacity $z_\sigma=\exp(\mu_\sigma/k_BT)$, with $\mu_\sigma$ the chemical potential, is determined by $n_\sigma=-{\rm Li}_{3/2}(z_\sigma)/\lambda^3_\sigma$ where $\lambda_\sigma=(2\pi/k_BTm_\sigma)^{1/2}$ is the thermal de Broglie wavelength.
 For the interaction energy we use Eq.\ (\ref{eInt}), with $E_+$ the energy of the repulsive polaron as given by the many-body result found in Sec.\ \ref{sec:ManyBody} (see Fig.\ \ref{fig:repPolaronEnergy}).
 The energy $E_+$ of the repulsive polaron in general depends on temperature,  but one can as a first approximation use the $T=0$ value at the low temperatures relevant for itinerant ferromagnetism.
  On the repulsive branch, the effective mass of the minority atoms is generally quite close to their bare mass, so we have in (\ref{energyPPmix})
  assumed $m_+^*= m_\down$.
   The phase diagram  can now be determined by applying the  Maxwell construction on $\fcal_{\rm mix}$.

\begin{figure}
\begin{center}
\includegraphics[width=0.5\columnwidth]{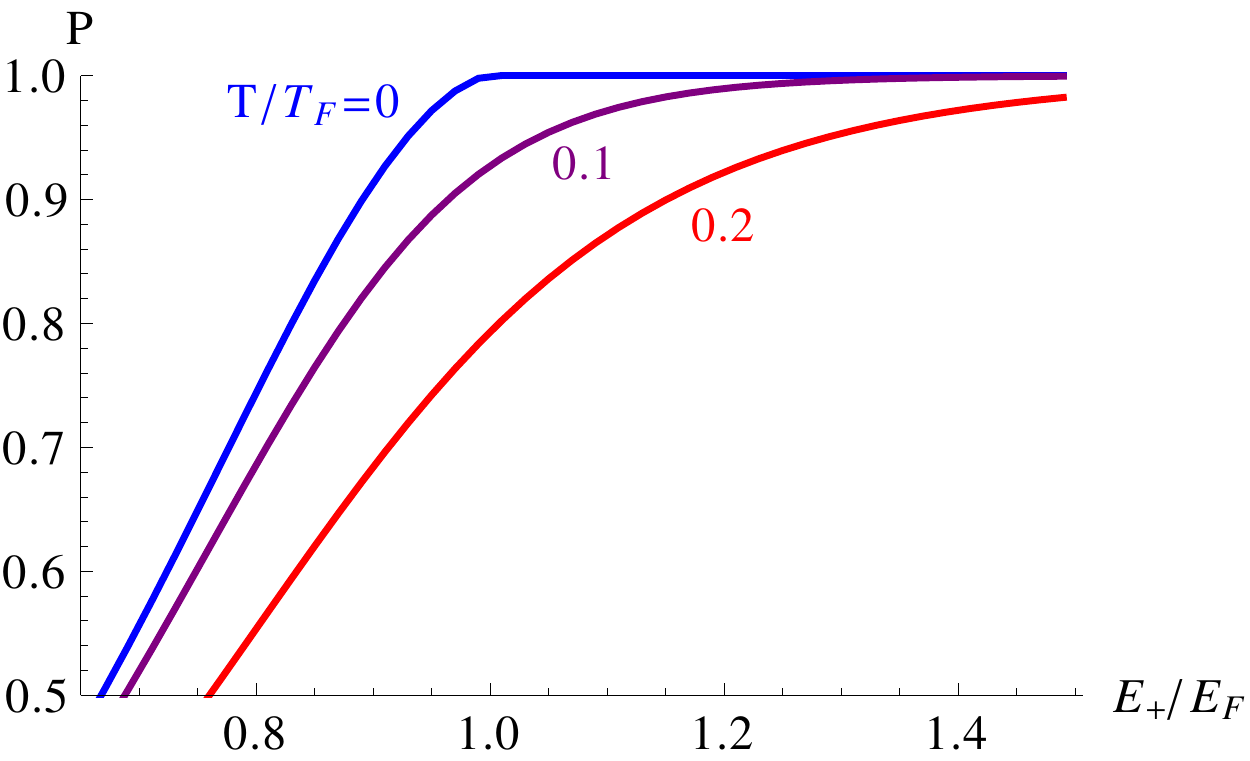}
\end{center}
\caption{Phase diagram showing the critical polarization $P$ for phase separation in terms of the polaron energy $E_+$ at various temperatures $T$. The gas is mixed above the lines, and phase separated below. Reprinted from Ref.~\cite{Massignan2013}. \copyright APS 2013.}
\label{fig:phaseDiagramVsEplus}
\end{figure}

In Fig.~\ref{fig:phaseDiagramVsEplus} we plot the phase diagram obtained from the Maxwell construction on the free energy (\ref{energyPPmix}), as a function of $E_+$ and polarisation $P$ for various temperatures~\cite{Massignan2013}.
We have only shown the diagram for $P\ge 1/2$ where this theory can be expected to be accurate and we have taken $m_\uparrow=m_\downarrow$.

For $T= 0$ and $P\to 1$, Fig.~\ref{fig:phaseDiagramVsEplus} shows
that the system phase separates when $E_+> \epsilon_F$.
 This reflects that the $\downarrow$-atoms cannot diffuse into a polaron state in the ideal gas of $\uparrow$-atoms if the polaron energy is higher than the
 Fermi energy \cite{Cui2010}. With decreasing polarization $P$,  phase separation occurs at a smaller polaron energy $E_+$ since the system can
 separate into two partially polarized phases, which reduces the kinetic energy cost.
 Conversely, we see that phase separation is  suppressed at higher temperatures due to the entropy of mixing.
 Note that the phase diagram in Fig.~\ref{fig:phaseDiagramVsEplus} is generic in the sense that  it is based only on the existence of well-defined and long-lived quasiparticles.

\begin{figure}
\begin{center}
\includegraphics[width=0.5 \columnwidth, clip]{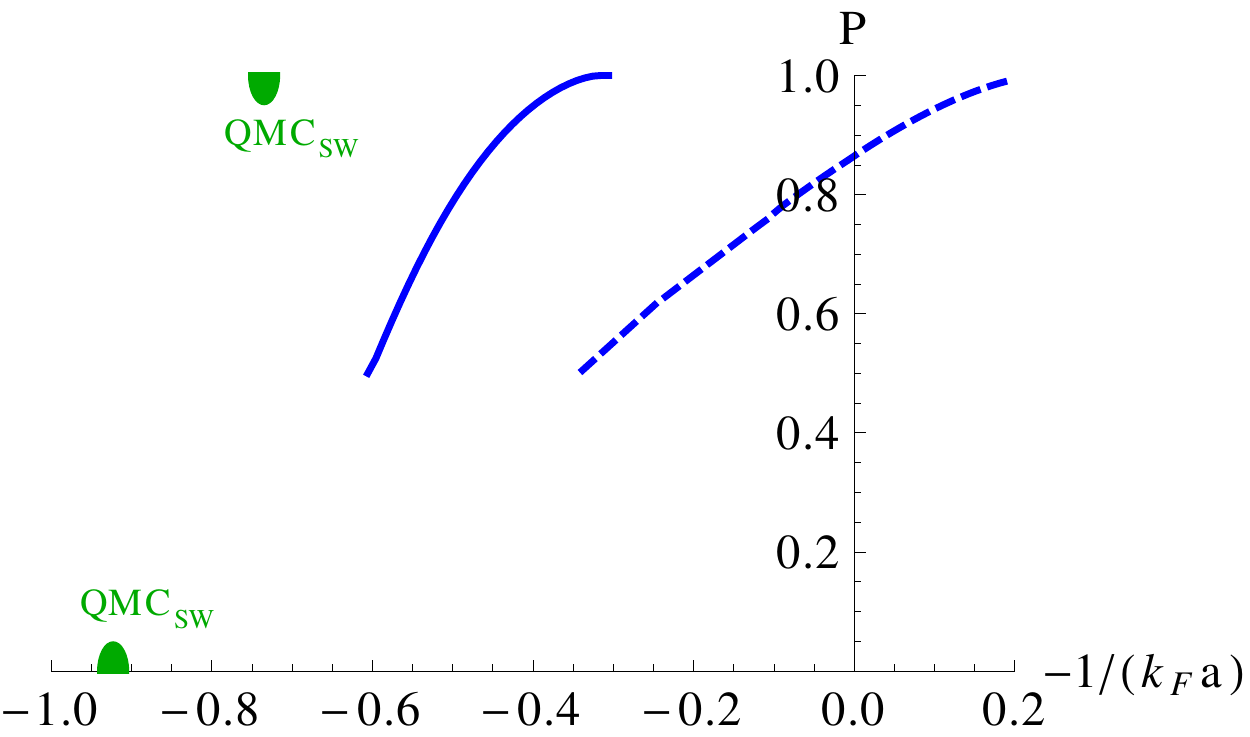}
\end{center}
\caption{Phase diagram of a repulsive Fermi mixture at zero temperature for two-body potentials with zero interaction range. The lines are obtained based on the polaron theory at zero temperature: results for equal masses with $k_FR^*=0$ (solid), and $k_FR^*=1$ (dashed). The gas is mixed above the lines, and phase separated below. The green markers indicate the IFM transition as found by Quantum Monte-Carlo calculations for a square well potential with negligible effective range ($k_FR^*\sim0$)~\cite{Pilati2010}.}
\label{fig:phaseDiagram_SW}
\end{figure}

From the theory described in section \ref{ManybodyModels}, one can calculate $E_+$ as a function of $k_Fa$ and therefore extract the phase diagram in terms of the physical interaction strength.
The result for $T=0$  and $m_\uparrow=m_\downarrow$ is shown in Fig.~\ref{fig:phaseDiagram_SW}
as the blue solid line for $k_FR^*=0$.
As in Fig.~\ref{fig:phaseDiagramVsEplus}, we have drawn the phase boundary lines only within the regime of validity of the polaron theory, i.e.~$P>1/2$.
We have furthermore terminated the lines towards the BCS side where the polaron Ansatz fails due to fast decay, as it will be discussed in more detail in section \ref{subsec:StabilityFerro}.

Due to the symmetry of the phase diagram for equal masses, the phase boundary  must cross the $P=0$ axis vertically.
This  restricts the range of possible extrapolations of the polaron theory from its range of validity to smaller values of $|P|$.
In particular, the extrapolation to $P=0$ predicts a critical value for ferromagnetism somewhere between the Stoner and the second order result.
The dashed line in Fig.~\ref{fig:phaseDiagram_SW} gives the phase boundary for
a resonance with $k_FR^*=1$. We observe that a non-zero effective range shifts the phase separated region toward the BCS-regime (the right side of the plot),
consistent with a similar shift of the
polaron energy discussed in section \ref{sec:ManyBody}.

As expected, the mass ratio plays an important role: In particular, when the minority atoms are heavy, $m_{\downarrow}/m_{\uparrow}>1$, itinerant ferromagnetism is favored since the Fermi pressure is reduced~\cite{vonKeyserlingk2011,Cui2013a}.
In the strongly polarized limit with a few heavy impurity atoms, the condition for phase separation at $T=0$ reads~\cite{Massignan2011}
\beq
E_+> \left(\frac{m_\up}{m_\down}\right)^{3/5}\epsilon_{F}.
\label{IFMcondition}
\eeq
which is smaller by a factor $(m_\up/m_\down)^{3/5}$ when compared with the equal mass case.

As mentioned in the previous Section, repulsive interactions induced by attractive potentials with short-ranges, such as square-wells (SW) or P\"oschl-Teller (PT) potentials, have also been studied via QMC techniques \cite{Conduit2009,Pilati2010,Chang2011}. In this case, fixed-node QMC can not be employed as the repulsive branch is an excited state of the many-body system. 
Instead, results based the variational Monte Carlo (VMC) method have been presented, using a trial wavefunction which is orthogonal by construction to the two-body ground state. Unfortunately, VMC is not as reliable as FN-QMC, since the orthogonality condition poses technical problems at strong coupling.
As a result, even though VMC calculations agree in predicting a transition to a partially polarized state at $k_{F\sigma}a=0.86$ for population balanced systems, they
 markedly disagree in the critical value for the transition to a fully polarized state, yielding respectively $k_{F\sigma}a=0.89$ \cite{Chang2011}, 0.92 \cite{Conduit2009}, and 
 1.08 \cite{Pilati2010}.

The QMC results of \cite{Pilati2010} for SW potentials at zero and full polarization are plotted in Fig.\ \ref{fig:PilatiPD} as red arrows, and in Fig.\ \ref{fig:phaseDiagram_SW} as the green markers ``QMC$_{SW}$". In the strongly-polarized regime $P\rightarrow 1$, we see that the critical coupling predicted by QMC-SW is smaller than the result predicted by the polaron theory. This is consistent with the fact that the repulsive polaron energy obtained by particle-hole expansion is lower than the one found by QMC, as we saw in Fig.\ \ref{fig:repPolaronEnergy}. The discrepancy between the two methods is by now unresolved, and its origin should be object of further studies. We finally mention two further calculations. The first, based on the Jastrow-Slater approximation \cite{Heiselberg2011}, found results compatible with variational QMC. The other, based on a modified Nozi\`eres-Schmitt-Rink theory which excludes contributions from the bound state~\cite{Shenoy2011}, instead found no diverging spin susceptibility, indicating the absence of a ferromagnetic transition at higher temperatures.

  \begin{figure}
    \begin{center}
\includegraphics[width=0.5\columnwidth]{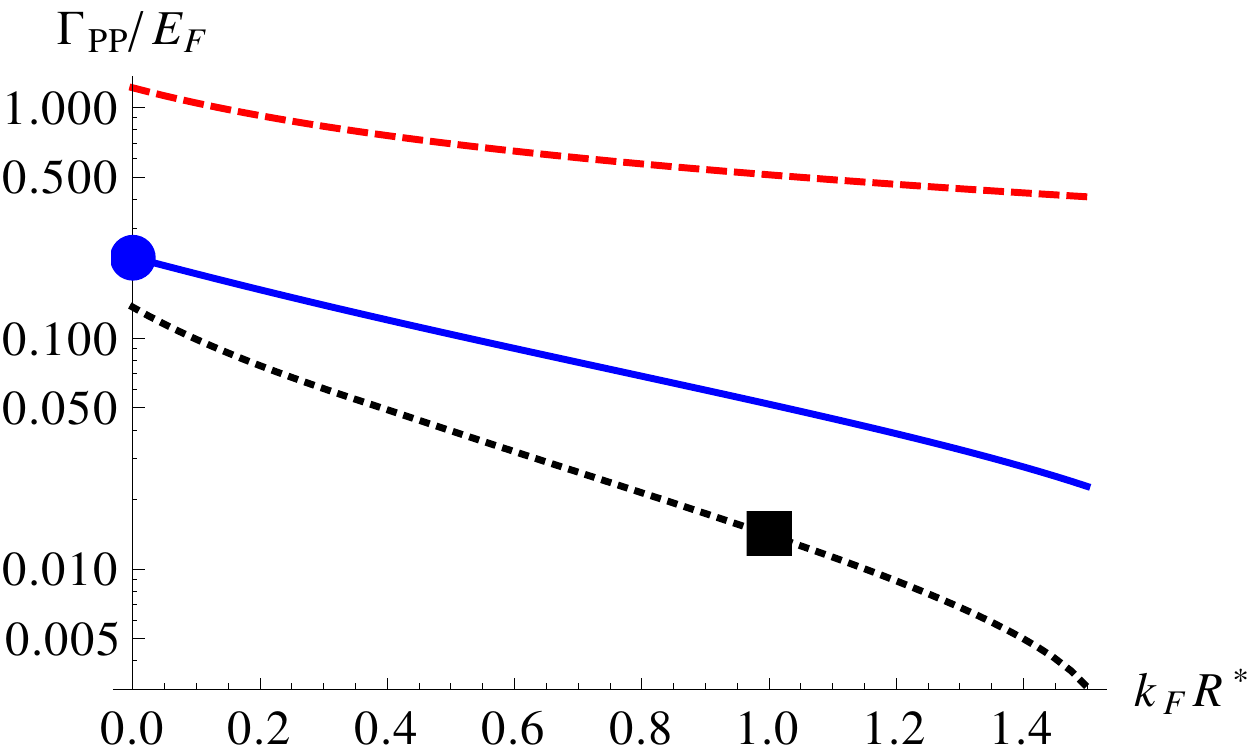}
  \end{center}
\caption{Decay rate $\Gamma_{\rm PP}$ of repulsive polarons at the critical coupling (\ref{IFMcondition}) for the ferromagnetic  transition with $P\rightarrow1$ and $T=0$,
 as a function of $R^*$. Lines are for mass ratios $m_2/m_1=$1 (solid), 40/6(dotted), and 6/40 (dashed). The circle and the square indicate the values relevant for the experimental conditions of, respectively, Refs.~\cite{Jo2009} and \cite{Kohstall2012}. Reprinted from \cite{Massignan2013}. \copyright APS 2013.}
\label{fig:decayRateAtTheTransition}
\end{figure}

\subsubsection{Stability of the upper branch}\label{subsec:StabilityFerro}
Since the upper branch is not the ground state of the system, it is important to examine
 its lifetime: is it long enough for the ferromagnetic correlations to be established and observed?

For a balanced system interacting via a broad resonance,
the decay rate $\Gamma_{\rm loss}$ of normal state atoms into the superfluid ground state was calculated from the imaginary part of the pair propagator~\cite{Pekker2011}.
Comparing the decay rate with the domain formation rate $\Gamma_{\rm df}$ of the ferromagnetic state as extracted from the imaginary part of the spin susceptibility, it was  found that the pairing instability always dominates the ferromagnetic instability. 
This calculation was later extended to the case of a non-zero polarisation $P$, with analogous conclusions~\cite{Sodemann2012}.
These findings seem therefore to rule out observing the ferromagnetic state for a system interacting via a broad resonance.

In the polarized case $n_2\ll n_1$, the question of decay can be reliably addressed using the polaron approach. The decay rate $\Gamma_{\rm PP}$ of the repulsive polaron  was calculated in section \ref{subsec:decay}. It
increases with increasing repulsion, and eventually the repulsive polaron  becomes ill-defined due to fast decay.
 For this reason, at strong coupling we have terminated the lines in
Fig.~\ref{fig:phaseDiagram_SW} when $\Gamma_{\rm PP}/\epsilon_F>0.25$.
 In Fig.\ \ref{fig:decayRateAtTheTransition} we plot the decay rate $\Gamma_{\rm PP}$ of the repulsive
 polaron calculated using the theory described in Sec.\ \ref{subsec:decay} at the critical coupling strength for phase separation at $T=0$ and $P\rightarrow 1$ for
 different mass ratios obtained from Eq. (\ref{IFMcondition}). Importantly, this figure shows that
  a resonance with  $k_F R^*\sim{\mathcal O}(1)$ gives rise to much longer lifetimes than a broad one with $k_F R^*\ll 1$. 
  A similar lifetime enhancement with large range was found for a balanced system in Ref.~\cite{Pekker2011a}. A large mass ratio $m_{\downarrow}/m_{\uparrow}$ 
  also decreases the decay rate significantly, compared to the case of equal masses.
For instance, for a mixture of a small number of $^{40}$K atoms in a gas of $^6$Li atoms, characterized by an effective  range $k_FR^*=1$, the polaron lifetime increases
by a factor $\sim 10$ at the critical coupling strength for phase separation when compared to the homonuclear, $k_FR^*=0$ case.  
This raises the prospect  of observing the ferromagnetic transition using atoms
interacting via a resonance which is not too broad. To examine this intriguing possibility further, as outlined above one should compare the decay time with the 
time to form ferromagnetic domains. In particular, one should calculate how the formation rate $\Gamma_{\rm df}$
depends on the effective range. Unfortunately, such a  calculation is not yet available. On the other hand, 
we expect the formation of domains to be mainly driven by the repulsive energy between the two spin components, which in 
 the highly-imbalanced limit is proportional to the polaron energy $E_+$.
 Since the decay rate in Fig.\ \ref{fig:decayRateAtTheTransition} is plotted for constant $E_+$, equal to the critical value for ferromagnetism given by Eq.\ (\ref{IFMcondition}), this figure indicates
  that, at constant $\Gamma_{\rm df}$, a narrow Feshbach resonance provides a large gain in the ratio $\Gamma_{\rm df}/\Gamma_{\rm loss}$, as compared to a broad one.

\subsubsection{Experiments with repulsive Fermi gases}
\label{subsec:IFMexperiments}
Experimentalists encountered strong repulsions already in the early days of the exploration of ultracold fermions, even though the first measurements were not directly targeted at the study of ferromagnetic phenomena.
A first characterization of elastic and inelastic scattering between weakly degenerate $^6$Li atoms on both sides of the  broad Feshbach resonance located around $830$G was reported in \cite{Jochim2002} and \cite{Dieckmann2002}.
Interestingly, the maximum of inelastic three-body decay was found to be significantly shifted on the BEC side from the resonance center.
Later \cite{Gupta2003} and \cite{Regal2004} performed inverse RF spectroscopy experiments on $^6$Li and $^{40}$K state mixtures respectively. Positive energy shifts up to a sizeable fraction of the Fermi energy were measured on the BEC side of the resonance~\cite{Regal2004}. A similar regime was reached in a $^6$Li mixture~\cite{Gupta2003}, but strong final state interactions complicated the interpretation of the experiment~\cite{Baym2007}.

\begin{figure}
\begin{center}
\includegraphics[width=0.5 \columnwidth, clip]{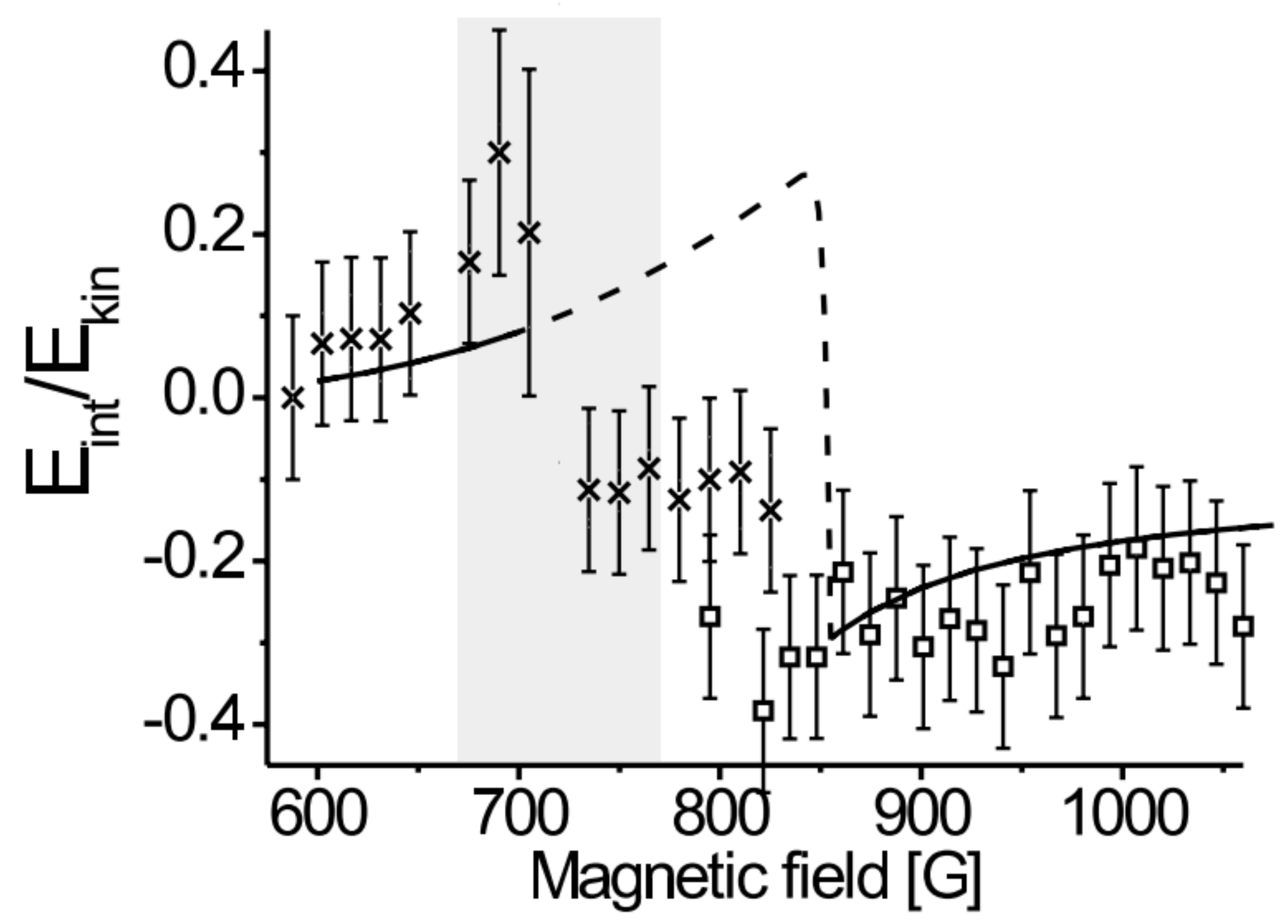}
\end{center}\caption{\textbf{Ratio of the interaction to kinetic energy} of a $^6$Li atom mixture across a Feshbach resonance. Measurements are taken by approaching the resonance from the BEC side (crosses) and from the BCS side (squares). The grey area indicates the region of strong losses, and the lines are a mean-field calculation. Reprinted from \cite{Bourdel2003}. \copyright APS 2003.}
\label{Bourdel}
\end{figure}

Of particular relevance here is the work of Ref.~\cite{Bourdel2003}, where the kinetic and the interaction energy of a weakly degenerate $^6$Li mixture were measured across the Feshbach resonance located at $B_0=830$G.
  Two sets of measurements were done: first, the mixture was prepared at a magnetic field $B_{i1}>B_0$
 on the BCS side of the Feshbach resonance, and then brought to a final field $B_f<B_{i1}$; second,
 the mixture was prepared at a magnetic field $B_{i2}<B_0$ on the BEC side with $a\sim 0^+$, and then brought to $B_f>B_{i2}$.
 In both cases the ramp duration was much longer than any trap period. The trap was then switched off and time-of-flight was performed, either keeping the magnetic field at   $B_f$, i.e. with interactions on, or with a magnetic field corresponding to a negligible interaction strength.
  In this way, the release energy $E_{\rm r}=E_{\rm kin}+E_{\rm int}$ and the kinetic energy $E_{\rm kin}$ could be measured, respectively.
Note that the possible presence of dimers influences the energy at $B_f$ whereas they should not contribute when there are no interactions during time-of-flight. Approaching the resonance  from the BEC side, the ratio of interaction to kinetic energy increased up to +0.3 for a magnetic field  corresponding to $k_F a \sim 1$,
whereas it  decreased and changed sign closer to the resonance, see Fig. \ref{Bourdel}.

Interestingly, these results have been interpreted in two different ways, both achieving rather good agreement. One interpretation is based on analyzing the energies and the stability of the attractive and repulsive branches of the Feshbach resonance using the virial expansion valid for high temperatures~\cite{Ho2004}, and it can be qualitatively captured with the toy model discussed in Sec.~\ref{subsec:toyModel}.
In this context, the experimental finding of \cite{Bourdel2003} when starting from the BEC side can be explained as it follows:
Initially, the system is on the repulsive branch while the attractive branch is empty. Approaching resonance, the
repulsion increases leading to an increased ratio of $E_{\rm int}/E_{\rm kin}$. At some interaction strength, three-body recombination from the repulsive to the
attractive branch becomes fast enough that a significant amount of dimers are formed during the ramp. Since the dimer binding energy is smaller than the trap depth, the  products of the recombination events remain trapped, creating a mixture of atoms and dimers.
As a consequence, the attractive branch  becomes macroscopically populated decreasing the interaction energy significantly, eventually making it negative. The sudden decrease in interaction energy at $B\simeq720$G (corresponding to $k_F a\sim 1$) in Fig.~\ref{Bourdel} is therefore due to the onset of a  macroscopic population of the attractive branch.
Similar conclusions were reached in Ref.~\cite{Zhang2011}.
Alternatively, the observed drop in the interaction energy can be interpreted as a signature of the formation of magnetic domains where the interaction
energy is strongly reduced~\cite{Duine2005}. These two very different interpretations of the same experiment, both  explaining the data
rather well, illustrate that
the behavior of fermionic mixtures in the upper branch of a Feshbach resonance is  difficult to analyze, since in general the system is a complicated non-equilibrium mixture of dimers and interacting atoms, with different instabilities competing simultaneously.

\begin{figure}
\begin{center}
\includegraphics[width=0.5\columnwidth, clip]{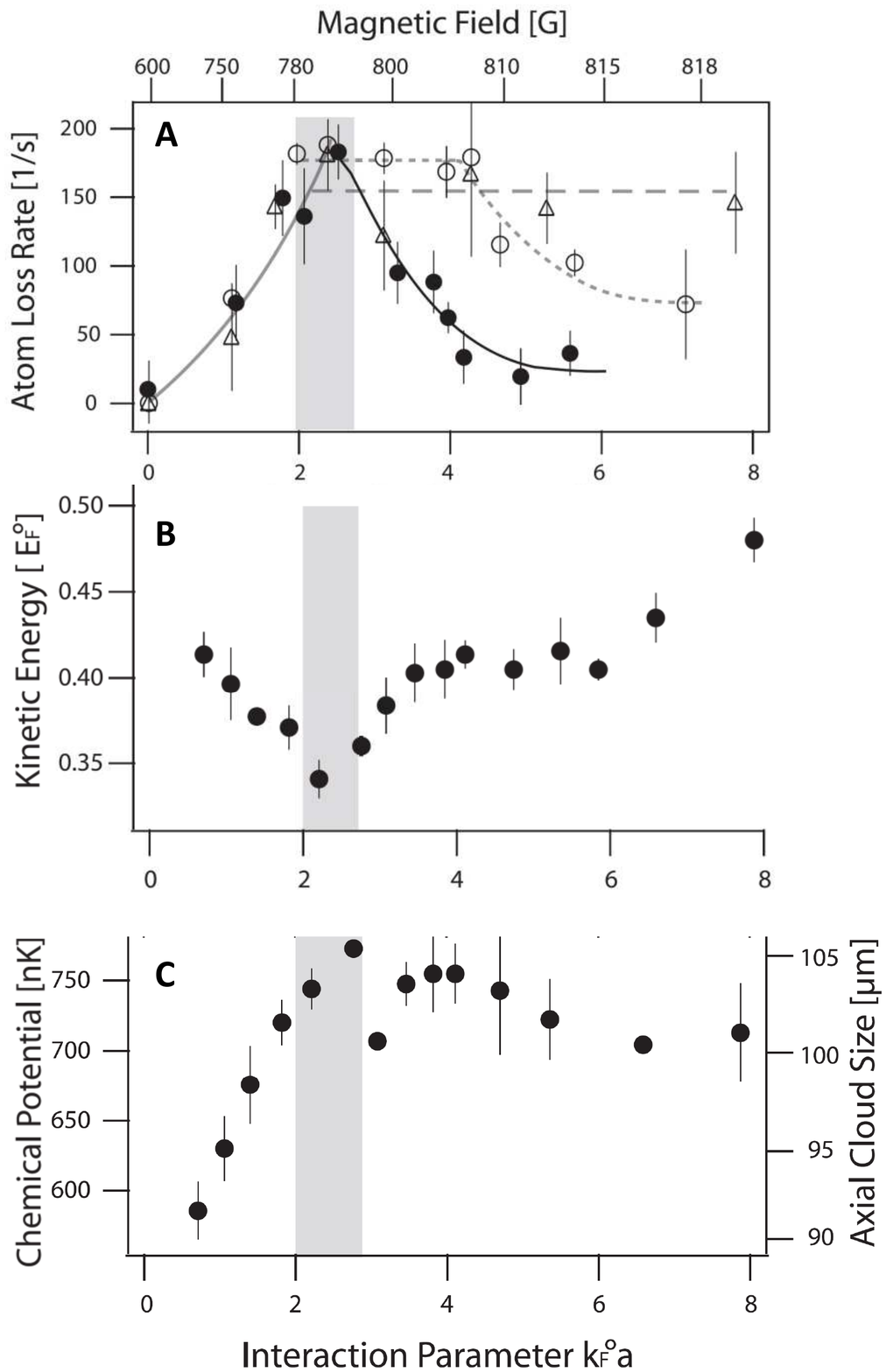}
\end{center}
\caption{\textbf{Properties of a degenerate $^6$Li Fermi mixture on the upper branch of a Feshbach resonance.} Filled dots are measurements at $T/T_F=0.12$.
 At a critical value of $k_F^0a\sim2.2$, the atom loss rate (A) shows a maximum, the kinetic energy (B) a minimum, while the chemical potential and the system size (C) saturate at a constant value. Reprinted from \cite{Jo2009}.
}
\label{Jo}
\end{figure}

More recently,  the group of W. Ketterle at MIT studied the upper branch using a $^6$Li mixture, specifically searching  for the ferromagnetic phase~\cite{Jo2009}.
Observables such as the loss rate, kinetic, interaction and potential energies were systematically investigated as a function of the  interaction
 for different temperatures.
As in \cite{Bourdel2003}, the system consisted of a 50-50 mixture of the two lowest internal states of $^6$Li atoms in the vicinity of the broad 830G resonance.
However, in these experiments one  could reach a much better degree of degeneracy with temperatures as low as $T/T_F=0.12$. To minimize the decay into the lower branch, the system was brought into the regime of strong repulsion with a fast ramp, not fully adiabatic with respect to the slowest trap period. Once the final field $B_f$ was reached,
 the inelastic collision rate was measured, and a sharp drop  was observed for $k_{F}^0 a\sim2.2$, see Fig. \ref{Jo}a. Here $k_F^0$ is the Fermi wavevector $k_{F\sigma}$ of the corresponding noninteracting gas  at the trap center.
As proposed by \cite{LeBlanc2009}, this was interpreted as the development of ferromagnetic domains leading to a decreased density overlap and
 therefore to a suppression of losses.
 The kinetic energy of the atoms furthermore showed a non-monotonic behavior (see Fig. \ref{Jo}b): it first decreased with increasing $k_F^0a$, but then it increased rapidly for $k_F^0a\gtrsim 2.2$.
  At the same time, \emph{in-situ} images measured  the axial size of the atom cloud. First, the size increased with the interaction strength, and it then saturated at a maximum value when $k_F^0a\gtrsim2.2$, as shown in Fig.~\ref{Jo}c.
 All these observations are in qualitative agreement with the expectations of the mean field model discussed in section \ref{sec:perturbation}, predicting a transition to a ferromagnetic phase, although some quantitative mismatch was observed. \textit{In-situ} differential phase contrast imaging \cite{Shin2006} was also used to  search for the
  formation of ferromagnetic domains. However, no domains were detected. Instead, the authors could estimate a maximum volume for the magnetic domains of about 5 $\mu m ^3$, containing about 50 atoms each.
The absence of larger polarized regions was ascribed to the short lifetime of the upper branch due to decay.

This notable experiment caused an intense theoretical debate. It was  proposed that, similarly to the case of \cite{Bourdel2003}, the MIT data may be interpreted not as a sign
of ferromagnetism, but rather in terms of the building-up of a short-range strongly correlated state \cite{Zhai2009,Cui2010}, or in terms of rapid decay to the
attractive branch \cite{Zhang2011,Pekker2011}.

 A later experiment by the same group at MIT~\cite{Sanner2012} provided further insights into the behavior this system.
For this second study, two major improvements were made to the experimental apparatus: the implementation of a speckle imaging technique, and the ability to perform rapid magnetic field sweeps. The first allows for a direct measurement of the spin fluctuations~\cite{Sanner2011}, which should be strongly enhanced in the vicinity of a ferromagnetic transition where polarized domains develop. The latter allows for a quenching of the system from the regime of weak interactions to the one of strong repulsion within a timescale comparable with the Fermi energy of the system.
A 50-50 mixture was prepared within a shallow optical trap at $T/T_F\sim 0.2$ in the region of weak repulsive interaction.
The mixture was then brought into the strongly repulsive regime via a fast, non-adiabatic ramp.
After a hold time, either the spin fluctuations or the atom and dimer population were measured.
An important observation was  the absence of any significant enhancement of spin fluctuations at any interaction strength: at most, an increase of fluctuations up to 1.6 times the value in the non-interacting sample was observed immediately after the quench to $k_{F\sigma}a\sim$2.3. For longer hold times, the spin fluctuations decreased.
This observation was interpreted as the absence of magnetic domain formation~\cite{Sanner2012}.
The time evolution of the number of atoms and dimers after a rapid quench to various values of the interaction strength was also measured in the experiment.  The evolution of atom losses and dimer production showed two different behaviors for strong repulsion.
For short timescales, the total population remained constant but a sizeable amount of atoms were converted into dimers: The production of dimers became progressively larger as the resonance was approached, with measured conversion rates up to one tenth of the Fermi energy for $k_{F\sigma}a>2$, which led to the creation of a considerable amount of dimers during the ramp. After this fast recombination dynamics, a quasi-equilibrium state was  reached at a new temperature determined by the heating associated with the dimer production. At long timescales a steady increase of the dimer fraction was observed, which was ascribed to the continuous evaporation which cools down the system and shifts the atom-dimer chemical equilibrium towards higher dimer fractions.
All in all, the authors concluded that the realization of the  Stoner scenario in ultracold Fermi gases with a short-range  interaction is ruled out \textit{tout court} at a broad resonance, due to the fast decay which occurs on a timescale on the order of the inverse Fermi energy~\cite{Sanner2012}.
However, as we will discuss in the next section, various routes towards the experimental realization of IFM in ultracold gases may still be envisaged.
Recently, the compressibility of a weakly repulsive fermionic system  has been investigated  by studying the quasi-equilibrium density profiles~\cite{Lee2012}.
 Within the interaction range explored, a small reduction of compressibility was observed which could be explained in terms of first order perturbation theory.
 The  pairing instability associated with the attractive branch prohibited the observation of second order effects, which become sizeable only for stronger repulsion~\cite{Lee1959}. 


\section{Conclusions and future perspectives}
\label{sec:Conclusions}
The study of the impurity problem offered us a simple yet surprisingly accurate description of the properties of a few minority atoms in a Fermi sea, in terms of well-defined quasiparticles whose properties may be  calculated using variational/diagrammatic calculations. 
The main properties of the  quasiparticles, such as the energy, the quasiparticle residue, the lifetime, and the effective mass, determined within such a framework, have been found in excellent agreement with the results of Monte Carlo calculations and with recent experimental findings. Importantly, the agreement holds even for strong interaction, for mass imbalanced fermion mixtures, and in the case of narrow resonances characterized by a large effective range.

 Interesting future directions in this field include the study of the dynamical properties of the impurities, and of the possible phases realized in presence of a large number of dressed quasiparticles. Of interest would also be to investigate new physics in the regime where the quasiparticle picture breaks down, such as in one spatial dimension, or for infinitely massive impurities. In particular, the possibility of having spin-dependent collisions with deeply trapped impurities would provide a direct and clean realization of the Kondo physics scenario.

In the regime of repulsive interactions, ultracold quantum mixtures provide an ideal set-up for the study of itinerant ferromagnetism, a phenomenon with many yet unanswered questions in condensed matter. Here we have reviewed  various existing theories and experiments which considered itinerant ferromagnetism in the context of ultracold atoms. We have
moreover seen that the repulsive polaron provides a fundamental building block
to write the equation of state of a strongly-imbalanced repulsive quantum mixture. Using this, we calculated the phase diagram for ferromagnetism,
and we addressed  the important issue of the lifetime of the repulsive branch.

 Still, the experimental quest for a repulsively interacting fermion mixture with a sufficiently long lifetime to allow for the development of a ferromagnetic instability remains
  an open issue.
As we detail below, a few interesting routes for future experiments can nonetheless be envisioned.

As discussed in Sec.~\ref{subsec:StabilityFerro}, systems with a large mass imbalance and close to narrow resonances exhibit a significantly slower decay rate.
 An intriguing possibility is therefore to use atoms interacting via a narrow Feshbach resonance
   to observe the ferromagnetic instability~\cite{Kohstall2012,Lee2012,Massignan2013}.
 Theoretical studies also suggest that the presence of an optical lattice  lowers the critical interaction strength for a transition to the ferromagnetic state, as well as it increases the stability towards three-body decay~\cite{Zhang2010,Ma2012,Pilati2013,Zintchenko2013}.

Few-atom systems may also be beneficial. Indeed, the spatial separation of the  mixture into two  domains needs a timescale, in an optimistic estimate neglecting
the slowing down of spin motion caused by collisions,
on the order of the trap period corresponding to the rate $\Gamma_{\rm df} \sim \omega$. If the loss rate is a sizeable fraction of the Fermi energy,
$\Gamma_{\rm loss}\sim \alpha E_F \sim \alpha \omega (6N)^{1/3}$, one can conclude that  ferromagnetic behavior can, in principle, be reached only in few particle systems, since
 $\Gamma_{\rm loss}/\Gamma_{sep}\sim \alpha (6 N)^{1/3}$. For $\alpha\sim0.1$, the two rates are comparable for $N \sim 200$ atoms
 which approaches the lower limit of a thermodynamic system.
 Indeed, experiments on few-fermions systems (with $N_\up+N_\down\sim 2, 3, \ldots, 8$) in one-dimensional traps \cite{Zurn2012,Wenz2013}
 benefit from increased lifetimes of the upper branch
  and could give interesting insights towards a deeper understanding of ferromagnetic phenomena,
  despite being far from the thermodynamic limit~\cite{Bugnion2013,Lindgren2013,Gharashi2013}.

Although one can rigorously exclude itinerant ferromagnetism in 1D for all values of $1/k_Fa>0$ \cite{Lieb1962}, it has recently been predicted that a ferromagnetic state could
 be reached in 1D by entering the so-called ``super-Tonks" regime \cite{Haller2009,Cui2013b,Volosniev2013}. The system is  prepared on the upper branch in the weakly-interacting BEC region and
then brought to the BCS side of the resonance by sweeping
 the magnetic field through the resonance stopping at $1/k_Fa\sim0^-$.
Interestingly, the system should be stable against losses in this regime since the true ground state consists of dimers with a very large binding energy  (of order the transverse trapping frequency $\omega_\perp$ confining the gas to a one dimensional configuration, which is large compared to all other energies in the problem). Ferromagnetism may therefore be probed without the stability problems encountered in higher dimensions.
  However, due to the orthogonality of the relevant wave functions and the integrability of the system it is not clear how the system can evolve from the mixed state to the ferromagnetic one. In Ref.\ \cite{Cui2013b} it has been proposed that even a tiny symmetry breaking field which destroys spin conservation may allow the transition towards the ferromagnetic state.

Another interesting possibility could be to reverse the experimental procedure and start with a system  created in a phase-separated state. The two atomic components
can be separated by means of a tight and thin optical barrier. One could then observe whether the system mixes or not after the barrier is lowered (either suddenly or adiabatically),
 as a function of interaction strength and temperature. Once a ferromagnetic state becomes thermodynamically favorable for strong  repulsion,
  one  expects  the diffusion time of one component into the other to slow down significantly. The advantage of this scheme is that the decay of atoms into
   the lower branch would occur, at least at the beginning, only at the interface between the two regions, which constitutes a negligible part of the whole system volume.

  An appealing way  to realize a repulsive two-fermion mixture \emph{on the lower branch} is to  start from an atomic
Bose-Fermi mixture prepared on the BCS side of the resonance. By sweeping across the resonance, with the bosons as the minority component, one could then
 convert the system into a mixture of two fermionic components, one being the bose-fermi dimers and the other being the excess unbound fermionic atoms.
Since the interaction between the unpaired atoms and the dimers is generally repulsive, and since the lifetime of such imbalanced mixtures is known to be relatively long close to resonance \cite{Zirbel2008}, this system could allow for a much longer time window  to observe the ferromagnetic instability. Whether it is possible to adiabatically prepare the
 dimer-fermi mixture without incurring in the mean-field collapse of the atomic system on the BCS side of the resonance \cite{Zaccanti2006} is at present unclear. Mass imbalance could be beneficial to overcome this problem.

Finally, we wish to mention that ultracold gases have been proposed as quantum simulators of many other magnetic systems \cite{Bloch2008,Esslinger2010,LewensteinSanperaAhufinger2012book}. Of particular interest is presently the possibility of investigating SU(N) magnetism with ultracold alkaline-earth atoms \cite{Gorshkov2010,Taie2012}, or the SU(3) Kondo effect by means of localized impurities in a spinless fermionic bath \cite{Nishida2013}. Recent experimental results include the quantum simulation of frustrated classical magnetism in triangular lattices \cite{Struck2011} and the realization of short range magnetic correlations induced by  strong super-exchange interactions \cite{Trotzky2008,Nascimbene2012,Greif2013}. Temperatures in optical lattices are unfortunately still too high for the emergence of true long-range magnetism, but rapid technical developments should allow us to enter this regime soon.

In conclusion, important progresses have been made in the past years, but many challenges are still open and awaiting for investigation. The great flexibility offered by ultracold gases gives us very interesting prospects to realize a variety of strongly-coupled systems, and we are confident that future studies will provide us important pieces of understanding in the fascinating field of quantum matter.

\ack
Special thanks to Dietrich Belitz, Stefano Giorgini, Rudolf Grimm, Jason Ho, Theodore Kirkpatrick, Maciej Lewenstein, Giovanni Modugno, Meera Parish, Dmitry Petrov, Sebastiano Pilati, Zhenhua Yu, and all the FeLiKx group in Innsbruck.
We wish to thank Boris Svistunov, Nikolay Prokof'ev, and Martin Zwierlein for kindly providing us their theoretical and experimental results.
We are particularly indebted to Jesper Levinsen and Leticia Tarruell for insightful discussions and a critical reading of the manuscript.

P.M. acknowledges funding from ERC AdG QUAGATUA,  EU IP SIQS, MICINN Project TOQATA (FIS2008-00784) and Fundaci\'o Cellex.
M.Z. was partially supported by the Lise Meitner programme of the Austrian FWF.
G.M.B. acknowledges support from the Carlsberg Foundation.
Finally, we wish to thank the ESF POLATOM network for financial support.

\section*{References}
\bibliography{polRev}
\end{document}